\documentclass[]{article}
\usepackage[titletoc,title]{appendix}
\usepackage{amsmath}
\usepackage{multirow}
\usepackage{mathtools}
\usepackage[tableposition=top]{caption}
\usepackage{amsfonts}
\usepackage{mathptmx}
\usepackage{graphicx}
\usepackage{epstopdf}
\usepackage{amssymb}
\usepackage{rotating}
\usepackage{natbib}
\usepackage{hyperref} % for displaying the \url correctly
\usepackage[
singlelinecheck=false % <-- important
]{caption}
\usepackage{subcaption}
\usepackage{authblk}
\newcommand*{\myovline}[1]{\overbracket[0.3pt][-1pt]{#1}}

\begin{document}

\title{Transitivity Correlation:  Measuring Network Transitivity as Comparative Quantity}
\author[1]{David Dekker\thanks{Special thanks go to the participants of the seminar series at the Center for Business Network Analyses (University of Greenwich)and at the CASOS seminar(Carnegie Mellon University). Correspondence should be sent to David Dekker, E-mail: david.dekker@gmail.com}}
\author[2]{David Krackhardt}
\author[3,4]{Tom A.B. Snijders}
\affil[1]{University of Greenwich}
\affil[2]{Carnegie Mellon University}
\affil[3]{University of Oxford}
\affil[4]{University of Groningen}

\label{firstpage}
\maketitle\vspace{\fill}

\begin{abstract}
This paper proposes that common measures for network transitivity, based on the enumeration of transitive triples, do not reflect the theoretical statements about transitivity they aim to describe. These statements are often formulated as comparative conditional probabilities, but these are not directly reflected by simple functions of enumerations. We think that a better approach is obtained by considering the linear regression coefficient of ties $i\,\rightarrow\,j$ on the number of two-paths $i\,\rightarrow\,k\rightarrow\,j$ for the $(n-2)$ possible intermediate nodes $k$. Two measures of transitivity based on correlation coefficients between the existence of a tie and the existence, or the number, of two-paths are developed, and called ``Transitivity Phi'' and ``Transitivity Correlation''. Some desirable properties for these measures are studied and compared to existing clustering coefficients, in both random (Erd\"os-Renyi) and in stylized networks (windmills). Furthermore, it is shown that under the condition of zero Transitivity Correlation, the total number of transitive triples is determined by four underlying features of any directed graph: size, density, reciprocity, and the covariance between indegrees and outdegrees. Also, it is demonstrated that plotting conditional probability of ties, given the number of two-paths, provides valuable insights into empirical regularities and irregularities of transitivity patterns.

\end{abstract}\pagebreak

\section{Introduction}
Transitivity is the qualitative aspect of the transitive triple configuration \citep[e.g.,][]{HollandLeinhardt1976} that occurs when there is a tie between an ordered pair of nodes $i$ and $j$, and there exist at least one node $k$, such that there are directed ties from $i$ to $k$, and from $k$ to $j$. 
Transitivity measures have been developed to measure the frequency or relative frequency of transitive triples in networks (e.g., Holland and Leinhardt, 1970; Frank, 1980). \citet{Newman2001} write: ``Clustering refers to the \textit{increased propensity} of pairs of people to be acquainted with one another if they have another acquaintance in common"[p.026118-12]\textit{(italics added)}. Corresponding to this, but using the more traditional term `transitivity' rather than `clustering', we refer to "network transitivity" as the network-level property that captures the increased propensity of pairs of nodes to be directly connected when they are connected through an intermediary node.

The motivation of this paper resides in the fact that although network transitivity has been quantified to describe theoretical processes, this quantification has been separate from those theories. As noted by Holland and Leinhardt (1976), many theoretical statements in sociology and psychology are framed in terms of transitive triple configurations. In fact, various strands of scientific literature (e.g., biology, sociology, social psychology, physics) assign an important role to this configuration. Yet, theories explicitly or implicitly using the concept of network transitivity refer to the frequency of transitive triples being higher \textit{relative to} some null situation, e.g. balance theory in psychology, epidemiology or diffusion in biology, medicine and marketing science. This means it is not enough to consider the frequency or relative frequency of transitive triples in a network, but a measure needs to capture the increase in frequency compared to a situation without transitivity. 
Here a set of measures is developed, presented and analyzed that can capture the increased propensity of transitive triples in networks.

It is important to distinguish this set of measures from other related concepts, especially, clustering. In social sciences, micro processes are considered at the basis of many macro social phenomena we can observe \citep[e.g.,][]{Coleman1990, Faust2010, Schelling1978}. A ubiquitous construct in studying consequences of such processes is the concept of clustering in social networks. That is, frequently such social structures are not merely random aggregations of ties, such as modeled in Erd\"os-Renyi digraphs or expander graphs. Rather, clustering is the extent to which there are regions in the network that have a higher density than there is between these regions. A micro process that has been foundational to clustering is transitivity. A macro level assessment of transitivity among triples is the global level property called \textit{network transitivity}. Clustering has been equated to network transitivity. For example, in their pivotal paper, \citet[p.026118-12]{Newman2001} state:``...clustering in social networks [is] also sometimes called network transitivity." However, the two concepts do differ. While most of the literature has focused on the global property of clustering, this paper explores the equally important property of network transitivity.

The ideas in this paper also build on the work of authors \citep[e.g.,][]{Holland1970,Wasserman1975,Feld1982,Faust2007} who have shown that the triad censuses of networks are highly associated with lower order properties (nodal and dyadic).
This implies that descriptive  network transitivity measures should control for the lower order properties. The purpose of this paper is to elaborate this.

In this paper, first a theoretical development of network transitivity measures is presented. Second, existing and new measures of network transitivity are defined and their properties described. Third, behavior of different measures is compared in examples (stylized, random, and empirically observed networks). Subsequently, findings and further research opportunities are discussed, and conclusions presented.

\section{Transitivity in Social Sciences}
\subsection{Theory Formalization and Generalization}
The importance of network substructures in theory construction is well exemplified by Heider's balance theory \citeyearpar{Heider1946,Heider1958}. The eminent psychologist, Fritz Heider, proposes balance theory as an explanation for the valence a person attributes to objects that she associates to another person. As such, balance theory demonstrates the importance of transitivity in formalizing social theory. Furthermore, \citet{CartwrightHarary1956} formalize and generalize Heider's balance theory from one triple to $n (n-1) (n-2)$ triples in digraphs. 

Such formalizations and generalizations allow us to make statements about substructures in whole systems, which are essential in allowing to make empirical descriptions of theoretical processes on a global level. The gap between local observations and the global nature of much social theory can be bridged ``... by examining local structural properties and determining whether they hold, on the average, across entire social systems." \citep{HollandLeinhardt1976}. Hence, average occurrence of local structures (or sub-structures) is considered an important descriptive statistic of whole systems as it allows to link social structure to global theoretical statements.

\subsection{Network Transitivity as a Comparative Quantity }
Transitivity is a property of ordered labeled 3-sub-graphs \citep{HollandLeinhardt1971,HollandLeinhardt1972} or triples. It thus not only plays a role as conceptual configuration in sociological theory, it is also an attractive statistical concept for network modeling. It's theoretical importance in much of social science stems from a Heiderian view that transitivity occurs in social interactions at a rate that is in excess of what we would expect by chance \citep{Davis1967}. This has led to statistical modeling of the frequency of transitive triples under a variety of null models \cite[e.g.,][]{HollandLeinhardt1971,Frank1988,Karlberg1999}.

Another view that shares the same Heiderian roots, yet deviates from the approach that focuses on enumerating the absolute or relative number of transitive triples can be derived from \citet{Newman2001}. They define network transitivity as: ``... the \textit{increased propensity} of pairs of people to be acquainted with one another if they have another acquaintance in common" [p.026118-12] \textit{(italics added)}. Here, the concept of network transitivity is not reflected by a mere average measure
of transitive triples, but rather an average increased propensity to form transitive triples. This definition suggests measuring an intrinsic comparative transitivity quality of a network. This contrasts, for example, an external comparison of transitive triple counts or ratios to an assumed network model.

In the literature, transitivity is measured usually as the ratio of transitive to potentially transitive triples \citep{HararyKommel1979,Frank1980,Karlberg1999} or as the average density of personal networks \citep{WattsStrogatz1998,Newman2001}.
These measures, based on relative frequencies, do not reveal much about an \textit{increased propensity}.

``Network transitivity" quantifies a statement about the \textit{comparative} frequency of transitive triples among relevant triples in the network. It reflects a structural hypothesis that refers to an elevated conditional probability of ties given at least one two-step between pairs of nodes. This is an intrinsic statement about the occurrence of non-vacuously transitive triples given two-step paths\footnote{This only covers part of the Heiderian view on balance, which considers also balance in vacuously transitive triples.}, or two-paths, compared to triples for which the condition does not hold. 

To define such a comparison, for a given observed digraph with $n$ nodes, we rely on two simple probability mechanisms.
In Section~\ref{s:centering} we use the probability distribution consisting of the random choice of an ordered triple  $(i,j,k)$ of vertices {($i\,\neq\,j,$}\ {$i\,\neq\,k$,}\ {$j\,\neq\,k$)} from the total of $n$ vertices. The probability distribution used in Section~\ref{S:correcting} is the random choice of a pair of vertices and will be further elaborated in that section.
For the first probability distribution we consider the triplet \citep{WassermanFaust1994} of tie variables $x_{ij},\, x_{ik},\, x_{kj}$, which are defined as the dichotomous (0/1) indicators of the existence of the ties $i\,\rightarrow\,j,\ i\,\rightarrow\,k$, and $k\,\rightarrow\,j$, respectively.
Probabilities under the random choice of an ordered triple $(i,j,k)$ will be denoted by $p$. The basic comparison is given by the difference between conditional probabilities of a tie, given a two-step path, and, a tie given no two-step path,
\begin{equation}\label{bal1prob}
p(x_{ij}=1 \mid x_{ik}x_{kj}=1) \, - \, p(x_{ij}=1 \mid x_{ik}x_{kj}=0),
\end{equation}
where a positive difference demonstrates an increased propensity towards transitivity. This difference reflects the most relevant alternative to the configuration central to the definition of \cite{Newman2001}, namely, the configuration where pairs of people are acquainted with one another if they have \textit{no} other acquaintance in common.

\section{Measurement of Transitivity}
In this section we define various measures that express the comparative frequency of transitive triples in a network.

\subsection{Difference in Conditional Probability and Centered Clustering Coefficient}
\label{s:centering}

For transitivity as a purely descriptive statistic, a common definition is the ratio of transitive to potentially transitive triples \cite[e.g.,][]{WassermanFaust1994}, as proposed by \citet{HararyKommel1979}:
\begin{align}\label{ClusCoef}
	C=\frac{\sum\limits_i\,\sum\limits_{j\neq i}x_{ij}\, \sum\limits_{k\neq i,j}x_{ik} x_{kj} }{\sum\limits_{i}\, \sum\limits_{j\neq i} \, \sum\limits_{k\neq i,j} x_{ik} x_{kj}}=
	\frac{\sum\limits_i\,\sum\limits_{j\neq i}\, \sum\limits_{k\neq i,j}x_{ij} x_{ik} x_{kj} }{\sum\limits_{i}\,\sum\limits_{j\neq i}\, \sum\limits_{k\neq i,j} x_{ik} x_{kj}} .
\end{align}
If the network is non-directed, this is equal to the well-known formula
\begin{equation}\label{Clus}
	C=\frac{3\times \text{number~of~triangles~in~the~graph}}{\text{number~of~connected~triples~of~vertices}},
\end{equation}
coined the clustering coefficient by \cite{Newman2001}.
This is equal to the first term in (\ref{bal1prob}),
\begin{equation}\label{eq:CasProb}
C = p(x_{ij}=1 \mid x_{ik}x_{kj}=1) .
\end{equation}
Comparing (\ref{bal1prob}) and (\ref{eq:CasProb}) immediately shows that (\ref{eq:CasProb}) is only a partial expression of theoretical statements about network transitivity, because it lacks a comparative aspect.

Another measure for transitivity is the clustering coefficient defined by  \citet{WattsStrogatz1998} as the mean of local transitivity around the nodes.
The version for digraphs is given by
\begin{align}\label{eq:locClusCoef}
LC=\myovline{LC_{i}}=\frac{1}{n}\sum\limits_i
\frac{\sum\limits_{j\neq i} \sum\limits_{k\neq i,j}x_{ij} x_{ik} x_{kj} }{OD_{i}(OD_{i}-1)} ,
\end{align}
where $OD_{i}=\sum_{h}x_{ih}$ is the outdegree of node $i$.
Just like (\ref{ClusCoef}), however, this is not a comparative measure.

To develop a measure that does have a comparative nature, just like (\ref{bal1prob}), we present the two by two table for the two random variables $x_{ij}$ and $x_{ik}x_{kj}$ under the probability distribution of randomly drawing a triple $(i,j,k)$. Here $x_{ij}$ indicates the existence of a direct tie between $i$ and $j$ and $x_{ik}x_{kj}$ indicates the existence of a two-path, i.e., an indirect connection.
\begin{table}[htb]
	\captionsetup{justification=centering, margin=1cm}
	\begin{center}
		\begin{tabular}{ l  c  c  c  r }
			& & \multicolumn{2}{c}{$x_{ik}x_{kj}$} & \\
			& & 1 & 0 & \\ \hline
			\multirow{2}{*}{$x_{ij}$}& 1 & $p_{11}$  & $p_{10}$ & $p_{1+}$\\
			& 0 & $p_{01}$  & $p_{00}$ & $p_{0+}$ \\ \hline
			& & $p_{+1}$ & $p_{+0}$ & $1$ \\
		\end{tabular}
		\caption{Transitivity Joint and Marginal Probabilities}\label{tab:2by2tab}
	\end{center}
\end{table}
The cells in Table~\ref{tab:2by2tab} contain joint probabilities, while the row and column sums give marginal probabilities, respectively. The joint probability's first index indicates whether $x_{ij}=1$ or 0, while the second index indicates whether $x_{ik}x_{kj}=1$ or 0. For example, $p_{11}$ is the joint probability of a tie between the pair $(i,j)$ and a two-step between this pair via $k$. In the marginal entries, a plus (+) indicates summing over both joint probabilities. For example, $p_{1+}$ is the marginal probability of a tie, which is the sum of the joint probabilities of a tie and a two-path, and a tie and no two-path.

Given that conditional probability is given by the ratio of joint probability and marginal probability, (\ref{bal1prob}) is equal to $ p_{11}/p_{+1}
~-~ p_{10}/(1-p_{+1})$. It is well known that this difference is the bivariate linear regression coefficient for dichotomous data \citep[see][for an excellent exposition]{FalkWell1997}. We use this expression to define (\ref{bal1prob}) as \textit{TPB} (Transitivity Phi Beta):
\begin{align}\label{eq:TPB}
	\textit{TPB}=  \frac{p_{11}}{p_{+1}}-\frac{p_{10}}{(1-p_{+1})} .
\end{align}

A bivariate regression coefficient is equal to the covariance between the two variables divided by the variance of the explanatory variable. This implies that another expression is
\begin{equation}\label{TPhiB}
\textit{TPB}=\frac{\text{cov}(x_{ij},x_{ik} x_{kj})}{\text{var}(x_{ik} x_{kj})},
\end{equation}
where again the variance and covariance are with respect to the probability distribution of randomly drawing a triple of nodes from the digraph.

This expression emphasizes that centering is the major difference with existing measures. The numerator in (\ref{TPhiB}), which we shall call Transitivity Covariance, is by definition a centered measure for the joint occurrence of ties and two-paths in an observed digraph. The measures in equations (\ref{ClusCoef}) and (\ref{eq:locClusCoef}), clearly are not centered. The centering is essential for the comparative nature of our measure for network transitivity.

A major advantage of centering is that it yields the value of~0 if there is no network transitivity in the sense that the existence of a two-path is not associated with the existence of a direct tie. For dichotomous variables a covariance of 0 is equivalent to independence; therefore, our transitivity measure \textit{TPB} is 0 if and only if, in case a triple $(i,j,k)$ is randomly drawn, the existence of the direct tie $i\,\rightarrow\, j$ is independent of the existence of the two-path $i\,\rightarrow\, k\,\rightarrow\, j$.
A direct expression for the Transitivity Covariance is the centered joint probability,
\begin{equation}\label{covPhi}
\text{cov}(x_{ij},x_{ik} x_{kj})=\frac{1}{n(n-1)(n-2)}  \sum_i\sum_{j\neq i} \sum_{k\neq i,j}x_{ij} x_{ik}x_{kj} - \myovline{x} \cdot \myovline{xx},
\end{equation}
where $\myovline{x}$ is the proportion of ties, or density in the digraph, and, $\myovline{xx}$ is the proportion of two-paths among all triples of nodes in the digraph.

Another measure can be obtained as the bivariate correlation coefficient instead of the regression coefficient. For this measure, bounded between $-1$ and $+1$, the Transitivity Covariance is divided not by the variance of the two-path indicator but by the product of the two standard deviations. As both variables are dichotomous, the Pearson product-moment correlation coefficient is also known as the $Phi$ coefficient \citep{FalkWell1997}. Here, we use the term ``Transitivity Phi",
\begin{equation}\label{TPhi}
	\textit{TPhi}= \frac{\text{cov}(x_{ij}, x_{ik} x_{kj})}{\sqrt{\text{var}(x_{ij})\text{var}(x_{ik} x_{kj})}}.
\end{equation}
The obvious further advantage of this measure is that it is bounded between $-1$ and~$+1$.

\subsection{Correcting for Two-path Autocorrelation}
\label{S:correcting}
The measures proposed in the preceding section do not take into account the multi-level issue that for each pair $(i,j)$ there are $n-2$ potential vertices $k$, which play a different role in the triplet than $i$ and $j$. The `clustering' of two-paths through specific $k$'s for a given $(i,j)$, which may be called the autocorrelation between different two-paths connecting the same pair $(i,j)$, is ignored.

Considering the set of all potential `third' vertices $k$ leads to interest in the relation between the total number of two-paths connecting $i$ and $j$, and the existence of a direct tie $i\,\rightarrow\,j$. Therefore we now turn to the second probability model which is defined as the random draw of an ordered pair $(i,j)$. To distinguish this from the model of the preceding section we indicate the other nodes by the letter $h$, distinguishing them from the single third node $k$ in the preceding section. Accordingly we define the Transitivity Correlation\footnote{This measure has been implemented in function \textit{gtrans} in the R-package 'sna' \cite[p.112]{Butts2016}.} by
\begin{equation}\label{TC}
\textit{TC}=\frac{\text{cov}\Big(x_{ij},\sum\limits_{h\neq i,j}x_{ih} x_{hj}\Big)}{\sqrt{\text{var}(x_{ij})\text{var}\big(\sum\limits_{h\neq i,j}x_{ih} x_{hj}\big)}} .
\end{equation}
The relation between \textit{TPhi} and \textit{TC} is derived in Appendix \ref{AppTPhiTC}) and indeed depends on the two-path autocorrelation.

The other measure, similar to \textit{TPB} in (\ref{TPhiB}), replaces variance in two-paths for ordered triples $(i,k,j)$ with the variance of the number of two-paths for ordered node pairs $(i,j)$. This is the bivariate regression coefficient of ties on the number of two-paths between ordered pairs $(i,j)$,

\begin{align}\label{eq:TB}
 TB = \frac{\text{cov}\Big(x_{ij},\sum\limits_{h\neq i,j}x_{ih} x_{hj}\Big)} {\text{var}\Big(\sum\limits_{h\neq i,j}x_{ih} x_{hj}\Big)} ,
\end{align}

This slope gives a linear approximation of the conditional probability of a tie, given the number of two-paths. As such it is more informative about the increased propensity towards transitivity than for example the clustering coefficient $C$ in (\ref{ClusCoef}), which gives an mean conditional probability over all two-path counts.

At this point it should be emphasized that the expected values, covariances, etc., referred to in this paper are those of ties between randomly chosen vertices in an observed network, not those of possible underlying random graph processes. A disadvantage of this is that the measures discussed above can not be used for statistical inference without non-trivial additional assumptions. What is subtracted in centering is not the expected value under a null model for networks. As shown in the next section, a necessary condition for $\textit{TC}=0$ and $\textit{TPhi}=0$ is that the number of two-paths is a specified function of $n$, density, mutuals, and the covariance between in- and outdegrees.

However, there are random graph processes that do generate an expected value of $\textit{TC}=0$ and $\textit{TPhi}=0$. For example, in Erd\"os-Renyi digraphs we have
\begin{align}\label{eq:Bernoulli_Graph_covariance}
E \Big\{ X_{ij} \sum_{h \neq i,j} X_{ih}X_{hj} \Big\}
={}&
E\{ X_{ij} \} \, E\Big\{ \sum_{h \neq i,j} X_{ih}X_{hj} \Big\} .
\end{align}
This does show that the expected value of the numerator in the covariance measures under the Erd\"os-Renyi digraph model is $0$. Since what is subtracted takes account of the indirect connections, this centering is more subtle than the null expected value under the Erd\"os-Renyi digraph.

Further, we note that \textit{TC} and \textit{TPhi} differ only in the denominator, i.e.,  the standardization. Therefore one way of studying the differences between these measures is to consider the digraphs for which \textit{TC} or \textit{TPhi} are $-1$ or~$+1$ if such digraphs exist.

Digraphs that are unions of complete sub-graphs, to which also isolated points may be added, are completely transitive in the sense that $C=1$. If all these sub-graphs have the same size, then also $\textit{TC}=\textit{TPhi}=1$. However, if the sub-graph sizes are different, then $\textit{TC}$ and  \textit{TPhi} are less than~1. 
			
\iffalse
Now, the advantage of \textit{TPhi} that it is the weighted product of the values in (\ref{bal1prob}) does transfer to \textit{TC}. As established in appendix \ref{AppTPhiTC}, if the ratio $(\textit{TPB}/\textit{TC})$ is larger than or equal to~1 we can derive that $\textit{TC} > 0$ is a necessary and sufficient condition for an increased propensity of triples in a digraph to be transitive. \textit{TC} hence has some major advantages over \textit{TPhi}.
\fi
	 	
\section{Behavior of Transitivity Covariance Measures}
In assessing the utility of these centered measures we look at the behavior in comparison to existing clustering coefficients. Much of it will depend on the properties of Transitivity Covariance, when we know it to be zero.

\subsection{Descriptive Mathematical Properties}
Empirical studies find that the frequency of triangles in a network is to a large extent accounted for by lower order network properties \citep[e.g.,][]{Faust2007}. If it is totally accounted for by lower order properties it may be expected that transitivity covariance is close to zero. Then it would be concluded that there is no ``increased" (or decreased) propensity toward transitivity. The mean number of transitive triples over all ordered pairs $(i,j)$ is given by
\begin{equation}\label{TC0}
	TT \, = \, \frac{1}{n(n-1)} \sum_{i=1}^n \,\, \sum_{\mathclap{\substack{j=1 \\j\neq i}}}^n \enskip \, \sum _{\mathclap{\substack{h=1 \\h\neq i,j}}}^n \enskip x_{ij}\, x_{ih}\, x_{hj} \ .
\end{equation}
For any digraph, the condition $\textit{TC}=0$ is equivalent to
\begin{equation}\label{TC0}
	TT \, = \,
\frac{\big(n  \,\, \text{cov}(OD,ID) \,+\, n \, d^2 \,-\, M\big)\, d }{n (n-1)}
\end{equation}
(for a proof see Appendix \ref{app:Properties TCov}).
Equation (\ref{TC0}) expresses the necessary and sufficient condition for a zero correlation between $x_{ij}$ and $x_{ik}x_{kj}$. Therefore, if (\ref{TC0}) holds, which is equivalent to $\textit{TC}=0$, no elevated or decreased propensity to transitivity may be said to exist in the network. The number of transitive triples then is determined by a function of four parameters: number of nodes $n$, density (number of ties), reciprocity (number of mutual ties, $M$), and covariance between the degree distributions, $\text{cov}(OD,ID)$.

The fact that $\textit{TC}=0$ implies a conditioning on $\text{cov}(OD,ID)$ relates to the observation of Feld and Elmore (1982), who observe that ``... inequality of popularity among individuals implies disproportionate frequencies of particular types of triads, including transitive triad types". They do not make clear how the ``increased propensity" of transitive triples depends on degree. Transitivity covariance does control for such popularity induced transitivity as it incorporates the covariance between in- and outdegree.

\subsection{Telling Problem: Don Quichot Measures and Windmills}
\begin{figure}
    \centering
    \includegraphics[scale=0.9]{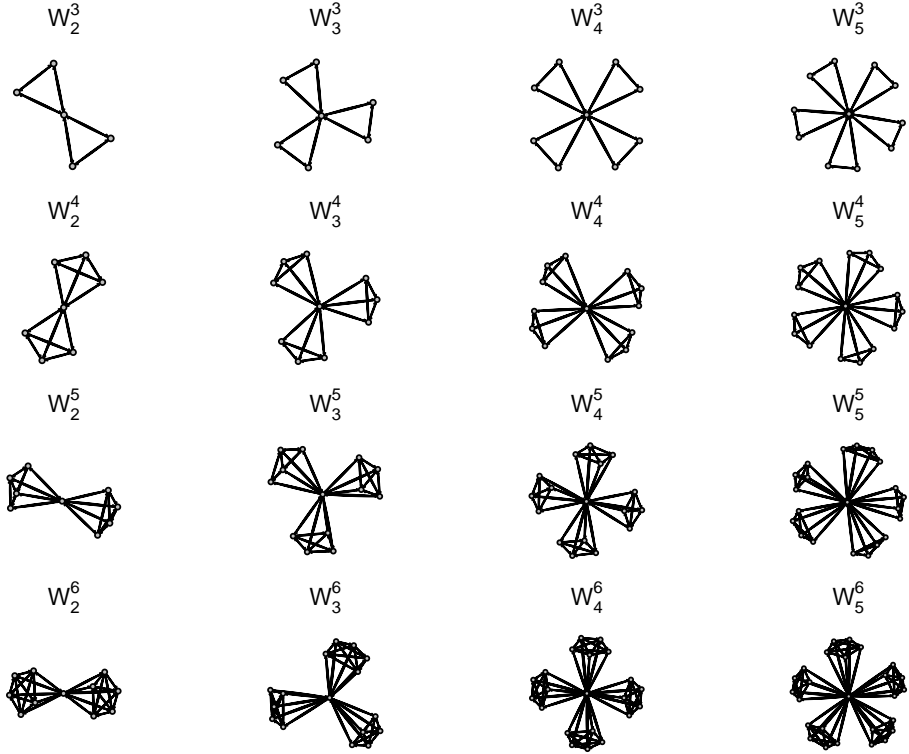}
    \caption{Windmills ($W^r_{m}$) with different values for $r$ and $m$}
    \label{fig:windmill}
\end{figure}

An example that illustrates the problems with different measures for an increased propensity of transitive triples in networks is given by structures named windmill graphs \citep[see for example][]{Jackson2008}. A windmill graph has one center node connected to all other nodes, while all other nodes are in 'wings', which are even sized cliques where all nodes are connected within wings, but not to any other node (except the center node). Windmills  $W_m^r$ are characterized by two parameters: the size of each wing $(r>2)$, and the number of wings $(m>1)$ (see for examples Fig.\ \ref{fig:windmill}).

In such graphs, there are either $1$ or $(r-2)$ two-paths between each pair of nodes, where the latter are always part of a transitive triple, while the former are not. Given this morphological constriction, windmills provide an experimental model that allows to vary the number of non-transitive two-paths and transitive triples as functions of $r$ and $m$. The number of two-paths in a windmill is given by
\begin{equation}\label{eq:WTS}
W_{TS}=m \, r\,(r-1)\,(r-2) \ + \ m\,(m-1)\,(r-1)^2,
\end{equation}
where the first part on the right-hand side is the number of transitive triples, while the second part gives the number of intransitive triples. The latter increases more strongly in $m$ as it is a quadratic polynomial, while the former is linear in $m$. The opposite holds for $r$ as the number of transitive two-paths increases cubically, and the intransitive triples quadratically, in $r$. Hence, this network model allows to manipulate the total degree of network transitivity.

To assess the behavior of network transitivity measures on the windmill model we express them in terms of $m$ and $r$. Table \ref{tab:WMcovmeasures} summarizes these expressions, as well as their behavior in the limit when either or both approach infinity. It is important to recall that the clustering coefficients in (\ref{ClusCoef}) and (\ref{eq:locClusCoef}) give the conditional probability of a transitive triple, and the \textit{mean} conditional probability of a transitive triple occurring in a neighborhood, respectively.   The contradictory effects of increases in $m$ and $r$ result in an undefined value for the clustering coefficient ($C$) in the bivariate limit, while it behaves as expected in the univariate limits. As $r$ increases it tends towards $1$, while it tends to $0$ with increasing $m$.

Similarly, the local clustering coefficient ($LC$), in the limits, reflects that apart for the single central node, all neighborhoods are cliques where all two-paths are transitive, so that it tends towards $1$. The contradictory outcome between $LC$ and $C$ as $m \rightarrow \infty$ was first noted in \citet[][p.36-37]{Jackson2008}.

In windmills, the transitivity covariance based measures, which are weighted functions of \textit{TPB} in (\ref{TPhiB}), illustrate another important distinction. For increasing wing size $r$, the difference in conditional probabilities (\textit{TPB}) still depends on the number of wings, $m$. On the other hand, when $m$ grows, the difference tends towards $0$ irrespective of $r$. The multivariate limit is undefined as it will depend on the
asymptotic ratio $m/r$.

\textit{TPhi} is restricted to [$-1,1$] as it is a correlation coefficient. In the limit in $r$ it becomes a decreasing function of $m$, and approaches $0$ for increasing $m$.
This is the correlation between $x_{ij}$ and $x_{ik} x_{kj}$ for a random triple $(i,j,k)$. For an increasing number of wings $m$ the conditional
probability of the two-path through a random $k$ between a given pair of nodes, i.e., $P\{ x_{ik} x_{kj} = 1 \| i,j \}$, tends to zero for all pairs
$(i,j)$; this implies that the correlation tends to 0.
The consideration of a random third node does not bring out the clustering pattern for windmills with many wings, and therefore this pattern yields approximately a zero correlation.

The covariance-based measures that weight on bases of the cumulative number of two-paths, \textit{TC} and $TB$, do signal this autocorrelation. First, \textit{TC} as a bounded measure on [$-1,1$] is a constant $1$, reflecting the perfect control for the morphological similarity of different size windmills. It indicates the perfect correlation that occurs in these structures, where the presence of a tie implies $(r-2)$ two-paths, while lack of a tie implies $1$ two-path --- the regularity that defines windmills.

In the limit $TB$ in windmills tends to $0$. The decline in the ratio of transitivity covariance (based on number of two-paths) and the variance of the number of two-paths is due to the fact that $Var(\sum_{h}x_{ih}x_{hj})$ is a factor $(r-3)$ larger than $Cov(x_{ij},\sum_{h}x_{ih}x_{hj})$.
This shows that for the value of \textit{TC} a direct interpretation is more clear than for \textit{TB}.

	\begin{sidewaystable}
		\begin{centering}
		\caption{Transitivity measures for Windmills}\label{tab:WMcovmeasures}
		  \begin{tabular}{ l  c c c c c  r }
		      Measure&$f(m,r)^*$&$\lim_{r \rightarrow \infty}f $ &$\lim_{m \rightarrow \infty}f$&$\lim_{(r,m) \rightarrow (\infty, \infty)}f$ & \\ \hline \hline\noalign{\smallskip}
		      	  \multicolumn{5}{l}{\textit{Clustering Coefficients:}} \\
		      $C$ & $\frac{r(r-2)}{r(r-2)+(m-1)(r-1)}$ & 1 & 0 & Undefined \\
		      $LC$ & $1-\frac{n-r}{n}\frac{1}{n-2}$ & 1 & 1 & 1 \\ \\ \hline \noalign{\smallskip}
		      \multicolumn{5}{l}{\textit{Composites for covariance based measures:}} \\
	          $Var(x_{ij})$ & $\frac{r(n-r)}{n^2}$ & $\frac{m-1}{m^2}$& 0 & 0\\ \\
		      $Var(x_{ik}x_{kj})$ & $\frac{((n-1)(n-2)-(r-1)(r-2))(n+r(r-3))}{n^2(n-2)^2}$& $\frac{m^2-1}{m^4}$ & 0 & 0\\  \\
		      $Var(\sum_{h}x_{ih}x_{hj})$ & $\frac{(m-1)r(r-1)(r-3)^2}{n^2}$&$\infty$ & 0 & Undefined\\ \\
		      $Cov(x_{ij},x_{ik}x_{kj})**$&$\frac{(m-1)r(r-1)(r-3)}{n^2(n-2)}$ & $\frac{m-1}{m^3}$& 0 & 0\\ \hline \noalign{\smallskip}
		      \multicolumn{5}{l}{\textit{Covariance based measures (triadic probability model):}} \\
		      \textit{TPhi}&$\frac{\sqrt(r)(r-3)}{\sqrt{(nr+r(r-3))(n+r(r-3))}}$ & $\frac{1}{\sqrt{m+1}}$& 0 & 0 \\ \\
		      \textit{TPB}&$\frac{(n-2)r(r-3)}{(n+r-3)(n+r(r-3))}$ &$\frac{m}{m+1}$ & 0 & Undefined \\ \hline \noalign{\smallskip}
		      \multicolumn{5}{l}{\textit{Covariance based measures (dyadic probability model):}} \\
		      \textit{TC}& 1 & 1 & 1 & 1 \\ \\
		      \textit{TB}& $\frac{1}{r-3}$& 0 & $\frac{1}{r-3}$ & 0 \\ \\ \hline \hline
		      \multicolumn{5}{l}{\^*$n=m(r-1)+1$, \^**Note that $\lim_{(r,m) \rightarrow (\infty,\infty)}(n-2)Cov(x_{ij},x_{ik}x_{kj})$ is undefined.}
	
	\end{tabular}
	\end{centering}
	\end{sidewaystable}

\subsection{Erd\"os-Renyi Random Digraphs }
The stylized example on windmills in the previous section shows that a family of morphological similar networks can produce measures that are undefined in the limit, while they may give ambiguous readings for small networks. This is not a desirable property. However, in practice other families of networks may be more important to consider.

\begin{figure}
    \centering
    \includegraphics[scale=0.5]{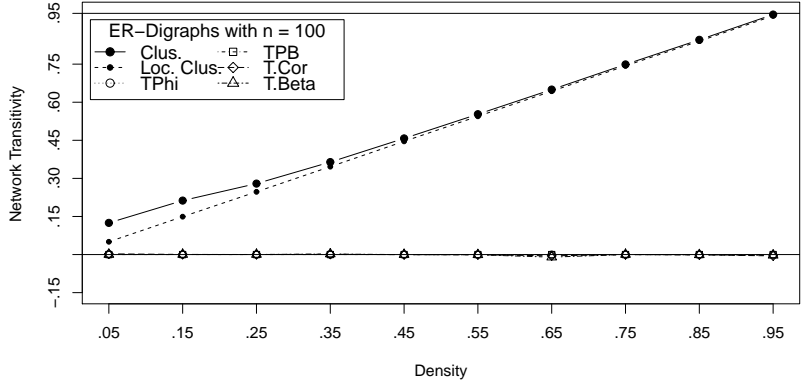}
    \caption{Means of Transitivity Covariance based measures and means of the clustering coefficients plotted as a function of density in random Erd\"os-Renyi digraphs ($n=100$). It becomes immediately clear that the clustering coefficients increase with density, while there certainly is no increased propensity of these random networks  to form transitive triples. The transitivity covariance-based measures remain stable around zero as density increases. Each point is a mean value of each of the measures based on $100$ random draws of Erd\"os-Renyi digraphs with given density.}
    \label{fig:TCvsCC}
\end{figure}

If there is known to be independence between ties and two-paths in a network there would not be expected any elevation or increased propensity in transitive triples, or network transitivity. Here we compare the behavior of different measures for Erd\"os-Renyi digraphs \citep{Erdos1959}. In these networks all ties are independent, and the probability for a tie is constant, determining the expected density. Hence, within this family of networks on average we do not expect to find any increased propensity for transitive triples to occur. Consequently, on average a transitivity measure should be independent of the density, in other words, control for the density.
Through simulations we first analyze the dependence of different measures on density. Figure~\ref{fig:TCvsCC} shows the results of these analyses for Erd\"os-Renyi digraphs. It shows that the covariance based measures are, as expected, independent of density; while $C$ and $LC$ are, respectively, linear and non-linear functions of density \citep[for the latter result see also][]{Newman2003b}.

We illustrate the measures by considering two networks harvested from our simulations of Erd\"os-Renyi digraphs. These networks exhibit transitivity and intransitivity, respectively. We specifically look at digraphs selected from simulations with $n=21$ and low mean degree, for graphical clarity.

\begin{figure}
	\captionsetup[subfigure]{justification=centering}
        \centering
        \begin{subfigure}[b]{0.75\textwidth}
                \centering
                \includegraphics[scale=0.75]{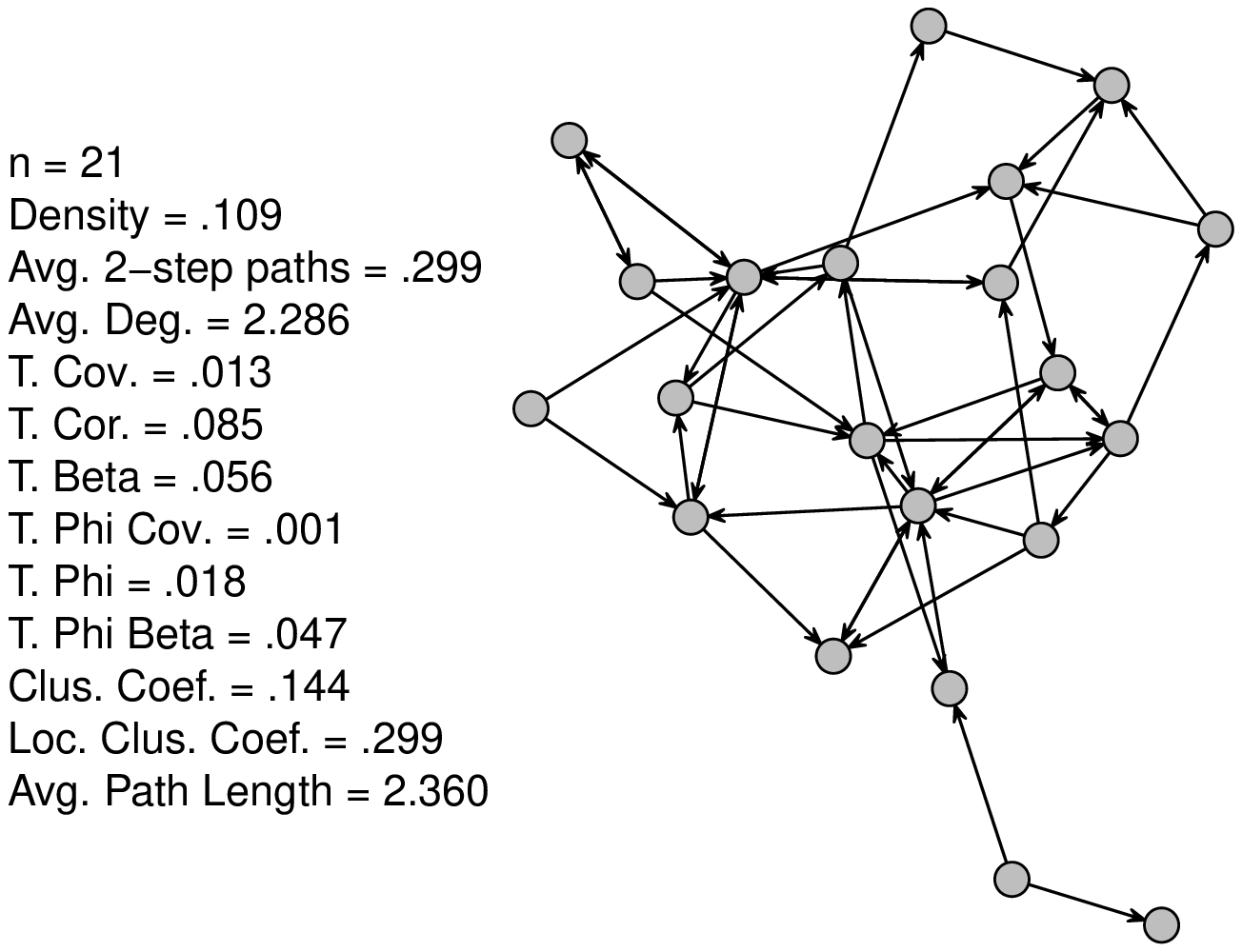}
                \caption{\\ \footnotesize{Positive value for Network Transitivity}}
                \label{fig:ERTC>C}
        \end{subfigure}
       \begin{subfigure}[b]{0.75\textwidth}
                \centering
               \includegraphics[scale=0.75]{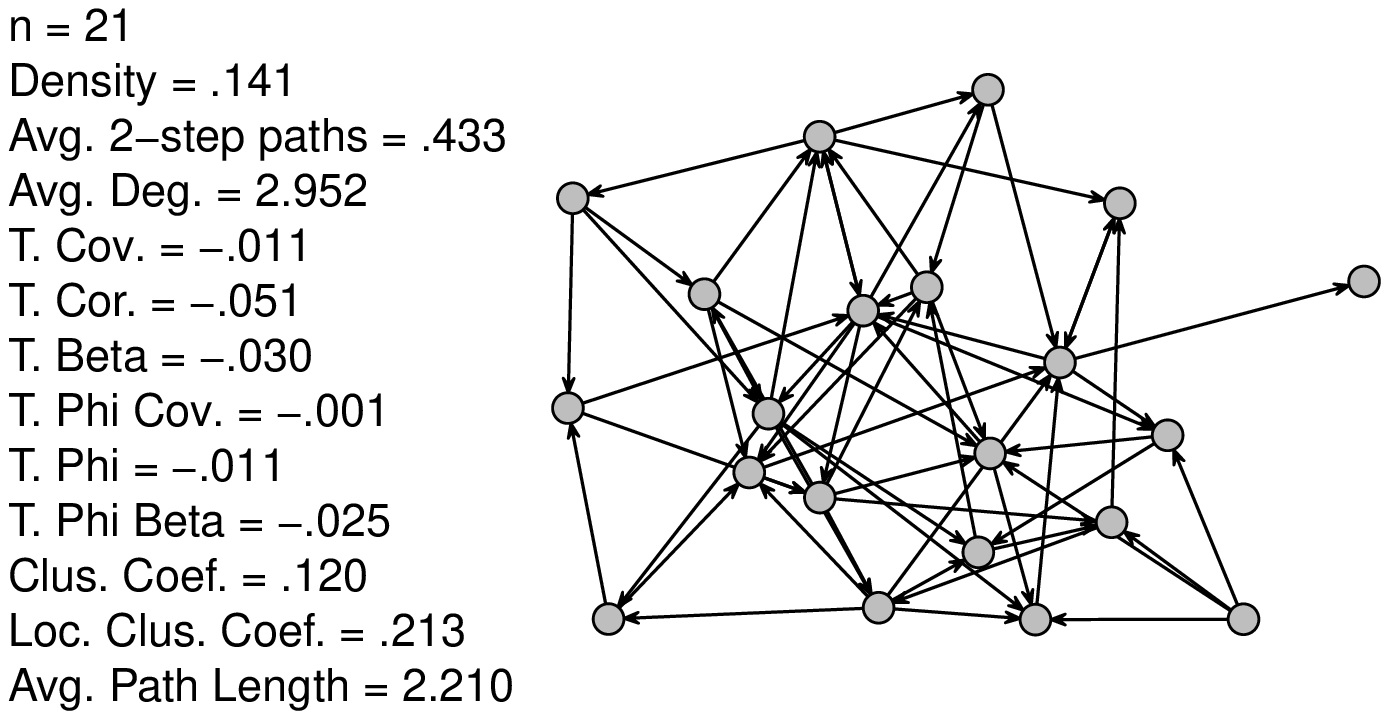}
                \caption{\\ \footnotesize{Negative value for Network Transitivity}}
                \label{fig:ERTC<C}
        \end{subfigure}
         \caption{Examples of positive and negative values for network transitivity in Erd\"os-Renyi digraphs ($n=21$)}\label{fig:TC&C instances}
\end{figure}

The cases we consider are depicted in Figure~(\ref{fig:TC&C instances}). The networks in Figures~(\ref{fig:ERTC>C}) and~(\ref{fig:ERTC<C}) are very similar on many properties, such as density, mean number of two−step paths, mean degree, and mean path length. Also the clustering coefficients are similar, at least they wouldn't lead to very different conclusions about the networks. Yet, the transitivity covariance measures indicate positive and negative values for network transitivity, respectively. \textit{TPB} shows that the probability of a transitive tie in Figure~(\ref{fig:ERTC>C}) is $4.7\%$ higher than the probability of a tie given there is no two-path, while the probability of transitive ties is $2.5\%$ lower in Figure~(\ref{fig:ERTC<C}). This difference in probabilities shows that in these cases, otherwise very similar, still there are opposite propensities toward the formation of transitive triples to form, with a large difference  of $(7.2 \%)$ in the contrasts of conditional probabilities.

\subsection{Observed Networks}\label{ObsNet}
The covariance based measures of transitivity can be interpreted as linear approximations of the relationship between direct ties and two-path ties. In particular, \textit{TB} is the linear regression coefficient of the tie indicator on the number of two-paths between the node pair. Graphical inspection of this relationship may provide insight about the appropriateness of the linearity assumption. Due to combinatorial restrictions the relationship may be highly non-linear, which can be directly assessed from a plot. Example datasets were obtained via public websites\footnote{Data via \cite{opsahldata}.\label{fnopsahldata}}\textsuperscript{,}\footnote{Data via  \cite{morenodata}.\label{fnmorenodata}}.

\begin{sidewaysfigure}
\captionsetup[subfigure]{justification=centering}
	\centering
	\begin{minipage}{1\textwidth}
		\begin{subfigure}[b]{.5\textwidth}
			\includegraphics[width=\textwidth]{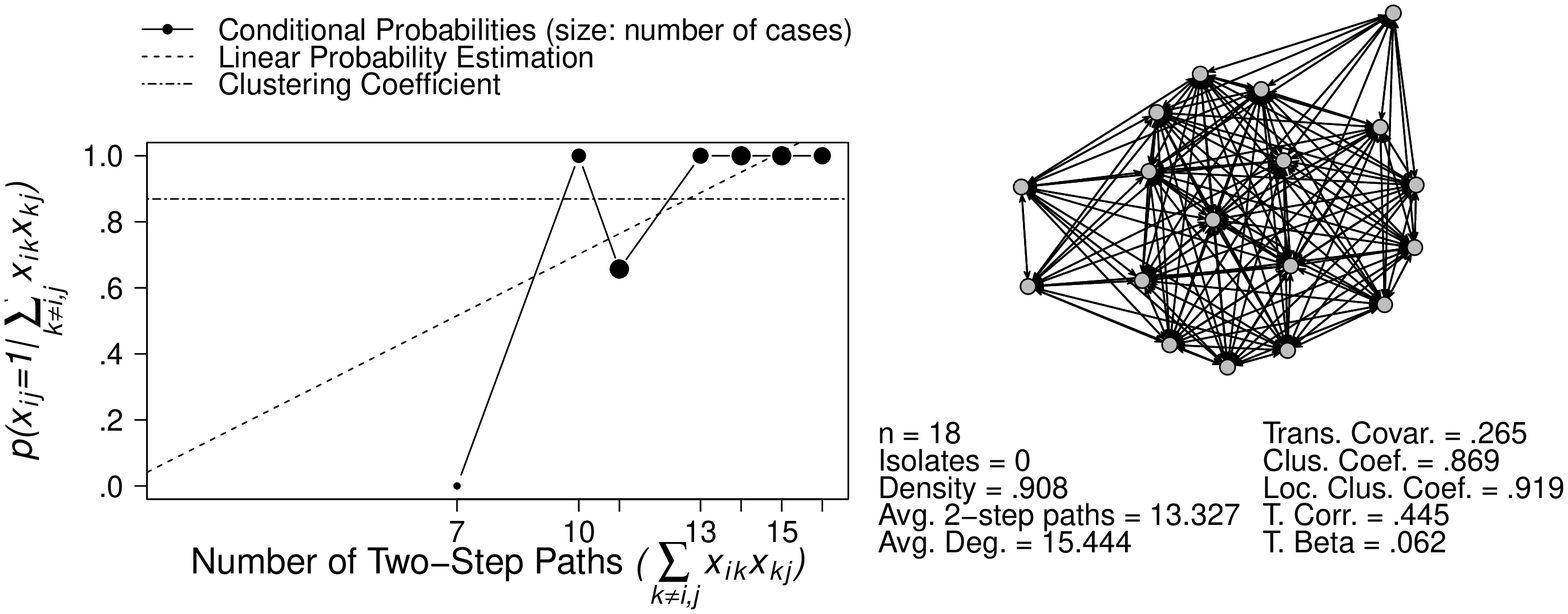}
			\caption{"Southern Women"-Data (\cite{DavisGardenerGardener1941}), cf. footnote \ref{fnopsahldata}. }\label{fig:souternwomen}
		\end{subfigure}\quad
		\begin{subfigure}[b]{.5\textwidth}

			\includegraphics[width=\textwidth]{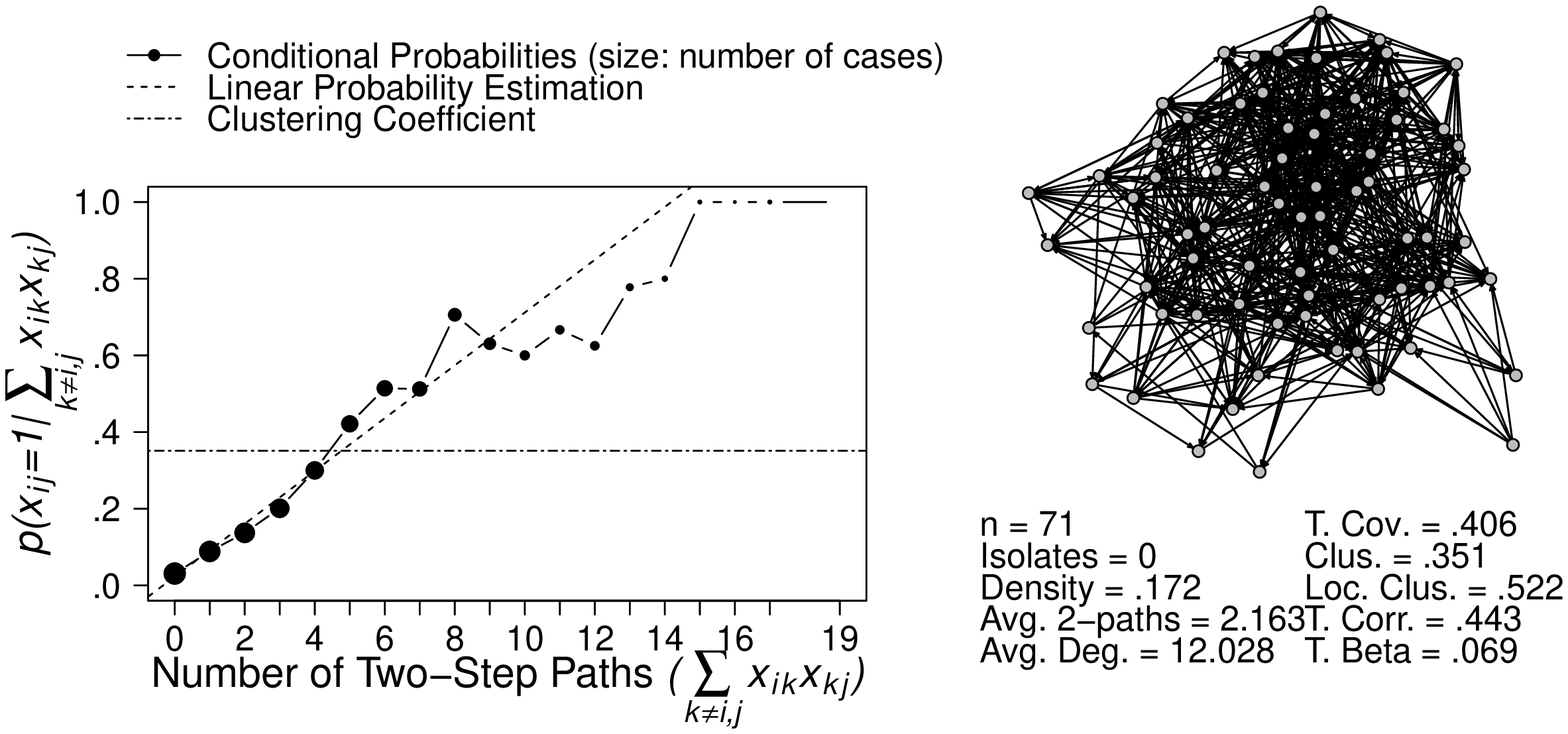}
			\caption{"Law Firm Friendships"-Data \citep{Lazega2001}, cf. footnote \ref{fnmorenodata}. }\label{fig:LazegaFriendship}
		\end{subfigure}
	\end{minipage}%

\medspace

	\begin{minipage}{1\textwidth}
		\begin{subfigure}[b]{.5\textwidth}
			\includegraphics[width=1\textwidth]{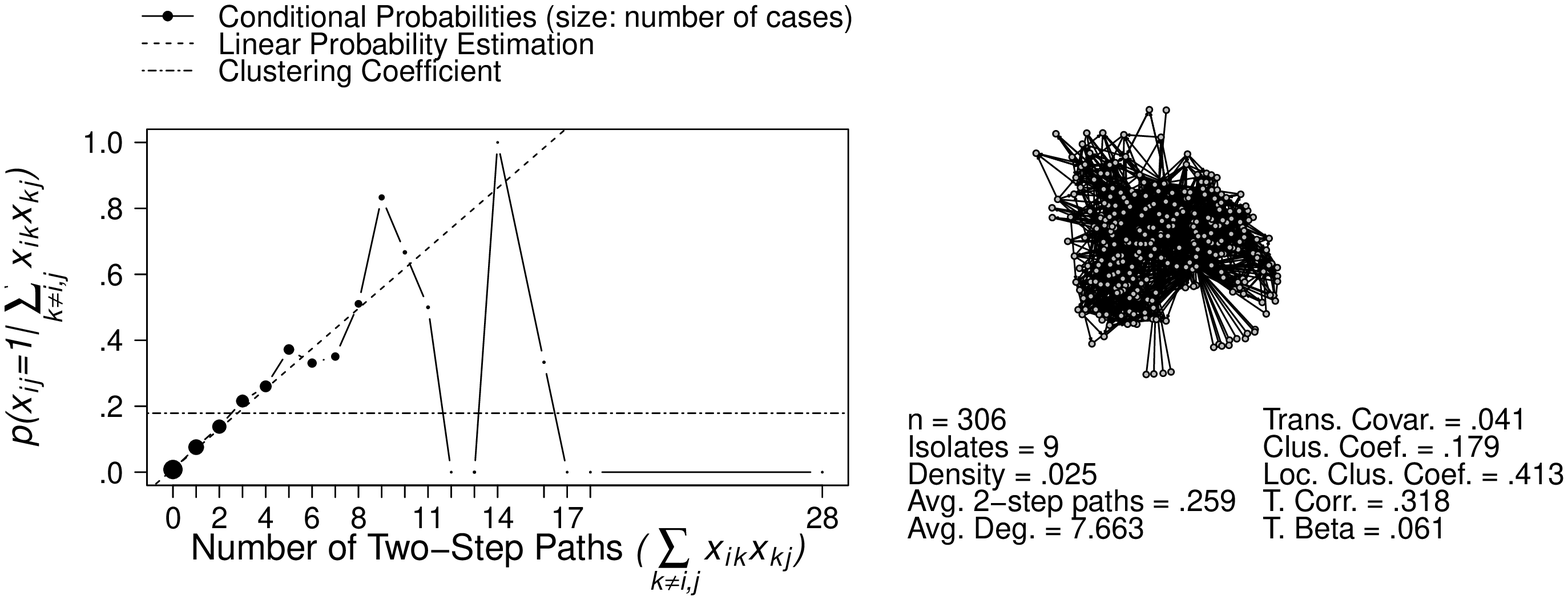}
			\caption{"Neural Network of Caenorhabditis elegans worm (C.elegans)"-Data \citep{WattsStrogatz1998}, Dichotomized, cf.  footnote \ref{fnopsahldata}.  }\label{fig:celegans}
		\end{subfigure}\quad
		\begin{subfigure}[b]{.5\textwidth}
			\includegraphics[width=1\textwidth]{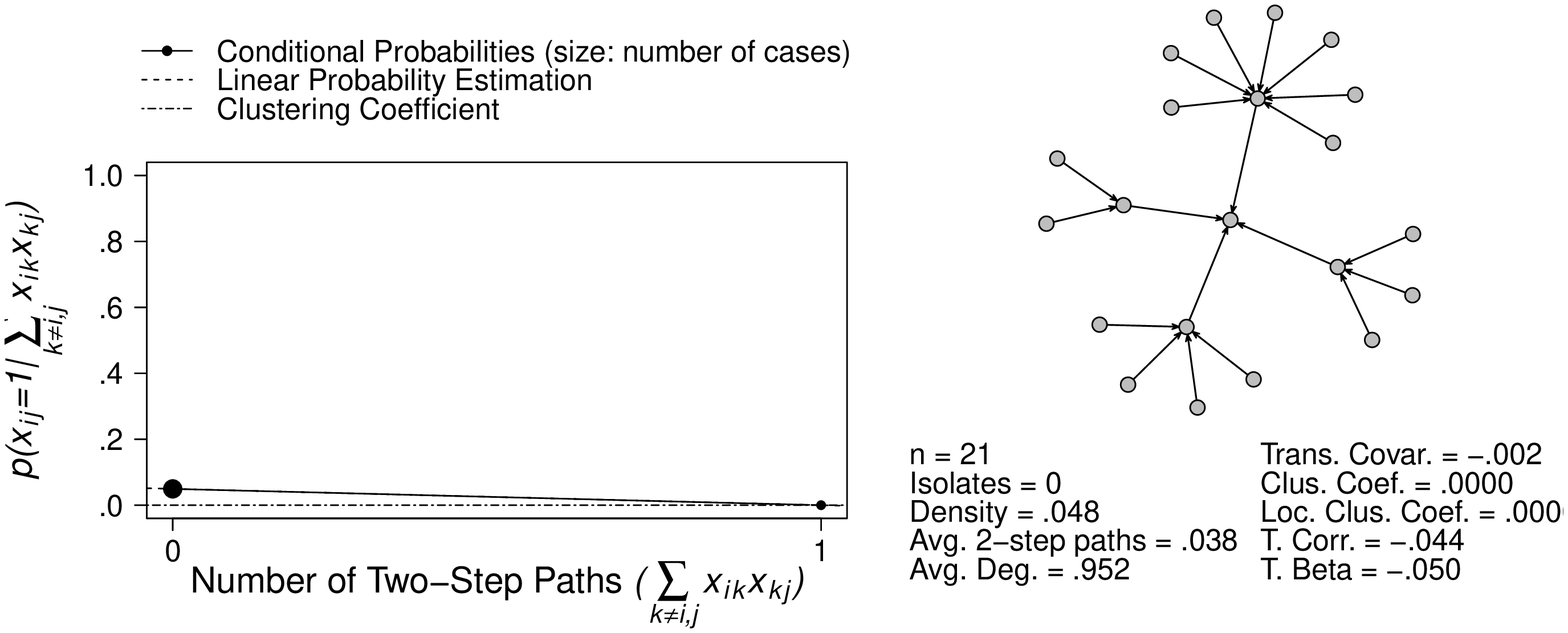}
			\caption{"High-Tech Managers Reports To"-Data \citep{KrackCSS87}, cf. footnote \ref{fnmorenodata}.  }\label{fig:reportsto}
		\end{subfigure}
	\end{minipage}

	\caption{A. Illustrating $10$ descriptive network statistics on 12 datasets. All reported graphs and descriptives are based on $x_{ij} \in \{0,1\} $. In case data are valued, they are dichotomized by the rule: $x_{ij}=1 ~ \forall ~ y_{ij}>0$ else $x_{ij}=0$, unless differently indicated.  On the vertical axis conditional probabilities for ties are shown, and dot sizes indicate the number of ordered pairs for each count of two-paths (horizontal axis). The horizontal line indicates the clustering coefficient ($C$) for that network. It can be interpreted as the weighted mean conditional probability over all groups of two-path counts. Third, the linear regression line between ties and number of ordered two-paths is shown. The slope of this line is given by $TB$ (T. Beta). The digraph plots show the networks without isolates.}

	\label{fig:empexa_A}
\end{sidewaysfigure}

\begin{sidewaysfigure}\ContinuedFloat
\captionsetup[subfigure]{justification=centering}
	\centering
	\begin{minipage}{1\textwidth}
		\begin{subfigure}[b]{.5\textwidth}
			\includegraphics[width=1\textwidth]{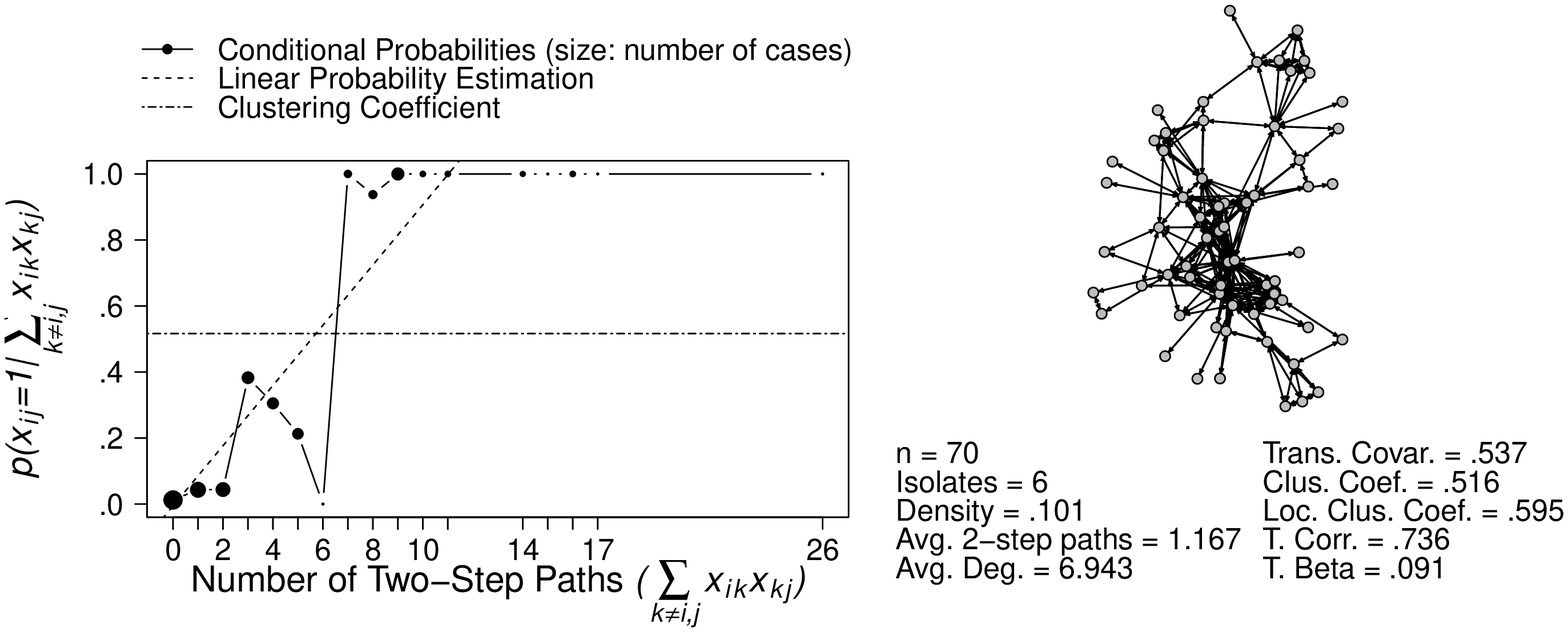}
			\caption{"Rodriguez Madrid Train Bombing"-Data \citep{Hayes2006}, Dichotomized, cf. footnote \ref{fnmorenodata}.   }\label{fig:Rodriguez}
		\end{subfigure}\quad
		\begin{subfigure}[b]{.5\textwidth}
			\includegraphics[width=1\textwidth]{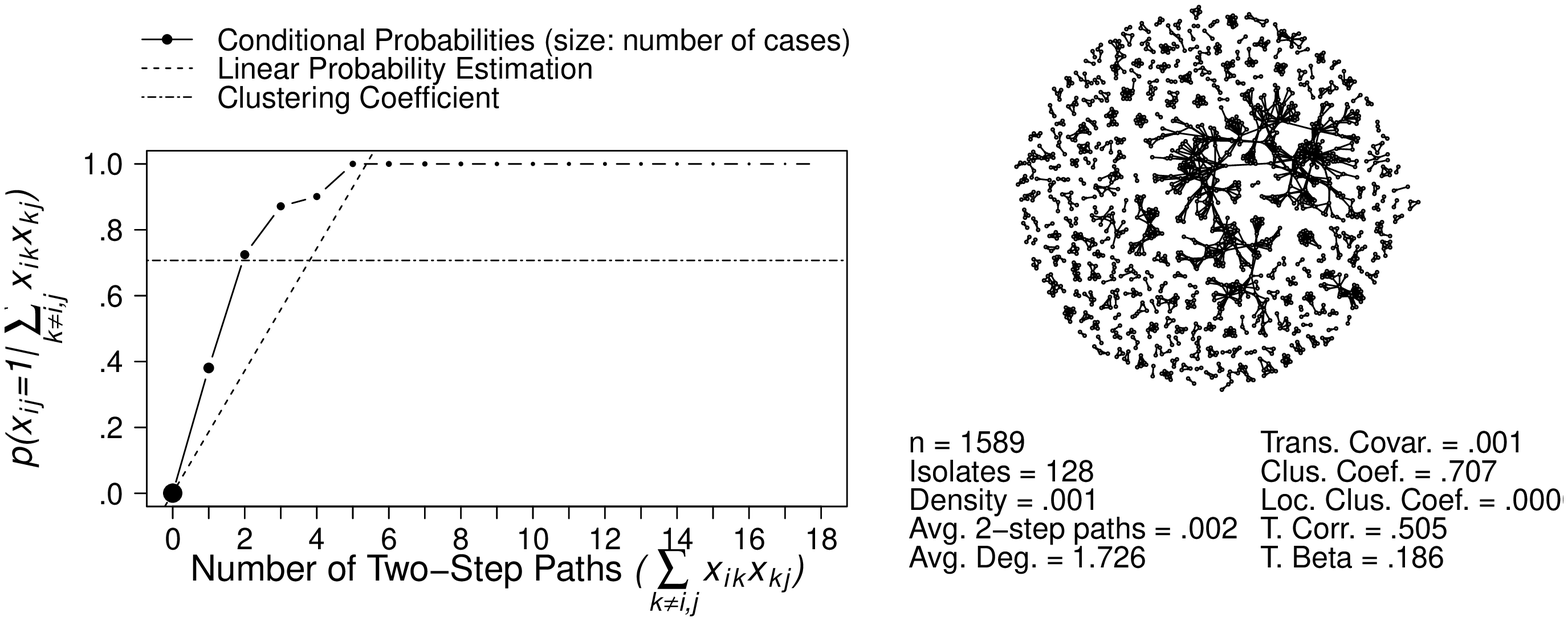}
			\caption{"Coauthors' Network"-Data \citep{Newman2006}, Dichotomized, cf. footnote \ref{fnmorenodata}.   }\label{fig:netsci}
		\end{subfigure}
	\end{minipage}

\medspace

	\begin{minipage}{1\textwidth}
		\begin{subfigure}[b]{.5\textwidth}
     		\includegraphics[width=1\textwidth]{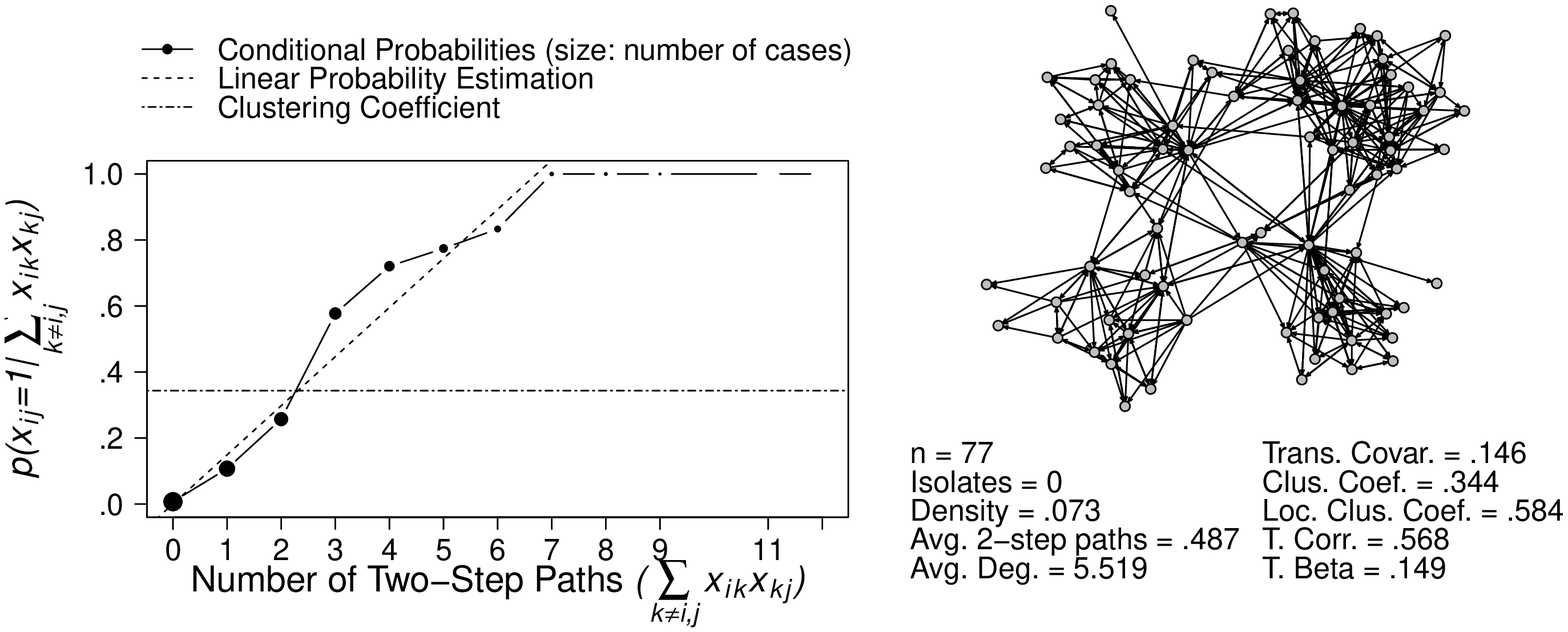}
			\caption{"R\&D Team in Manufacturing Company"-Data \citep{CrossParker2004}, Dichotomized $x_{ij}=1 ~ \forall ~ y_{ij}>4$, cf. footnote \ref{fnopsahldata}.   }\label{fig:RDadvice}
    	\end{subfigure}\quad
		\begin{subfigure}[b]{.5\textwidth}
		    \includegraphics[width=1\textwidth]{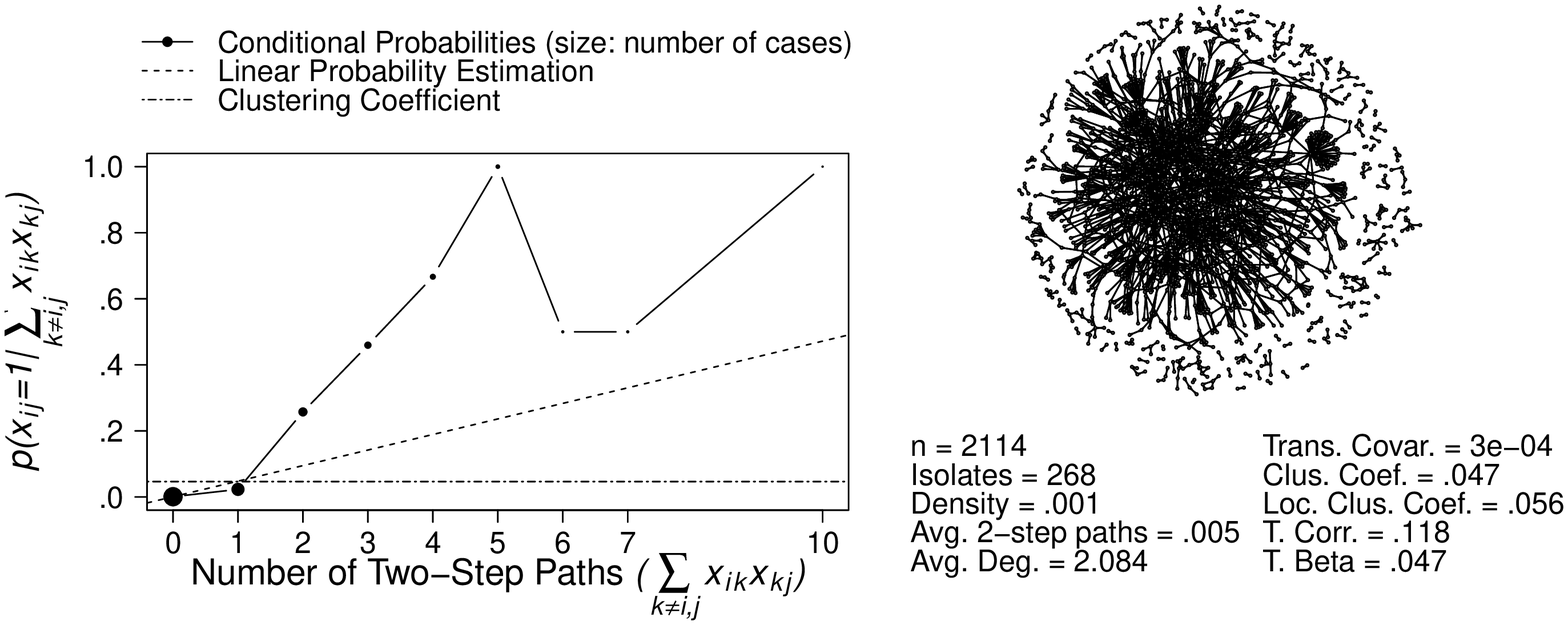}
			\caption{"Protein-Protein Interaction"-Data (\citep{Jeongetal2001}), cf. footnote \ref{fnmorenodata}.   }\label{fig:proteins}
	    \end{subfigure}
	\end{minipage}

	\caption{B}

	\label{fig:empexa_B}
\end{sidewaysfigure}

\begin{sidewaysfigure}\ContinuedFloat
\captionsetup[subfigure]{justification=centering}
	\centering

\begin{minipage}{1\textwidth}
	\begin{subfigure}[b]{.5\textwidth}
		\includegraphics[width=1\textwidth]{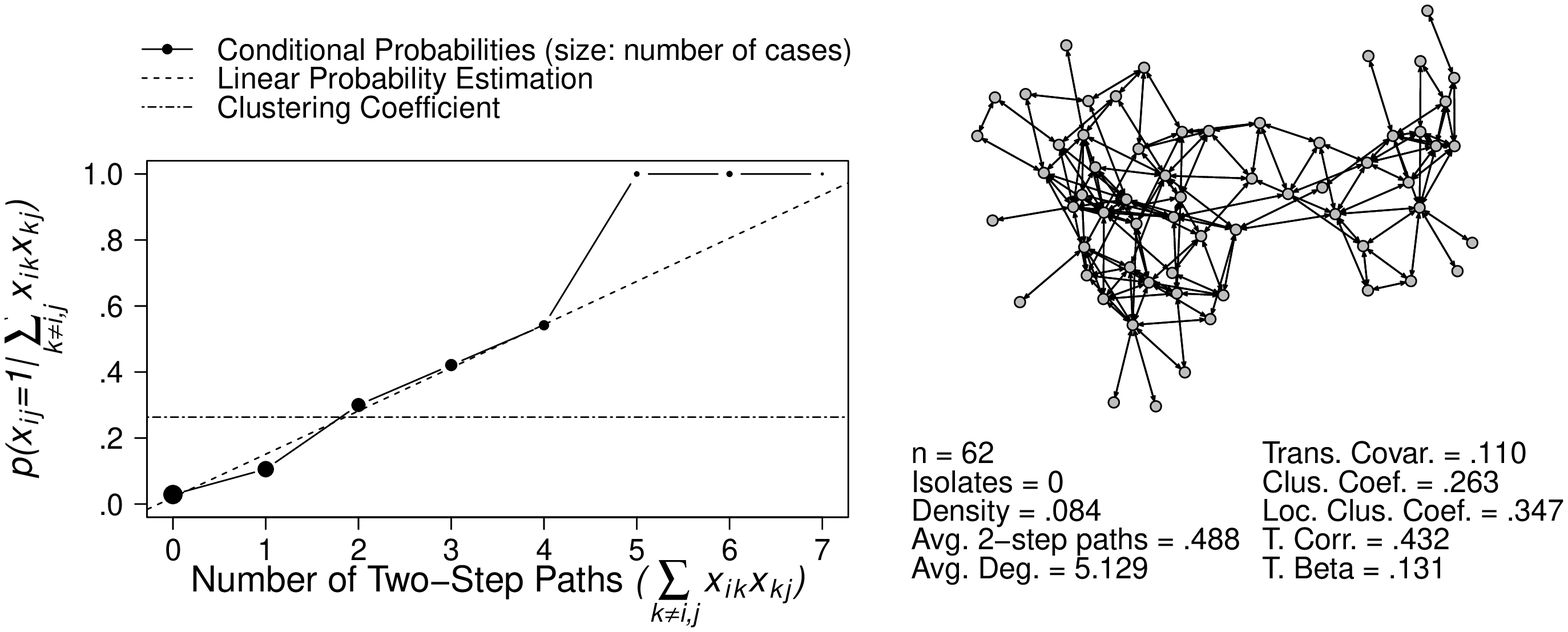}
			\caption{"Dolpins Frequently Associating"-Data \citep{Lusseau2003}, cf. footnote \ref{fnmorenodata}.   }\label{fig:dolphins}
	\end{subfigure}\quad
	\begin{subfigure}[b]{.5\textwidth}
	\includegraphics[width=1\textwidth]{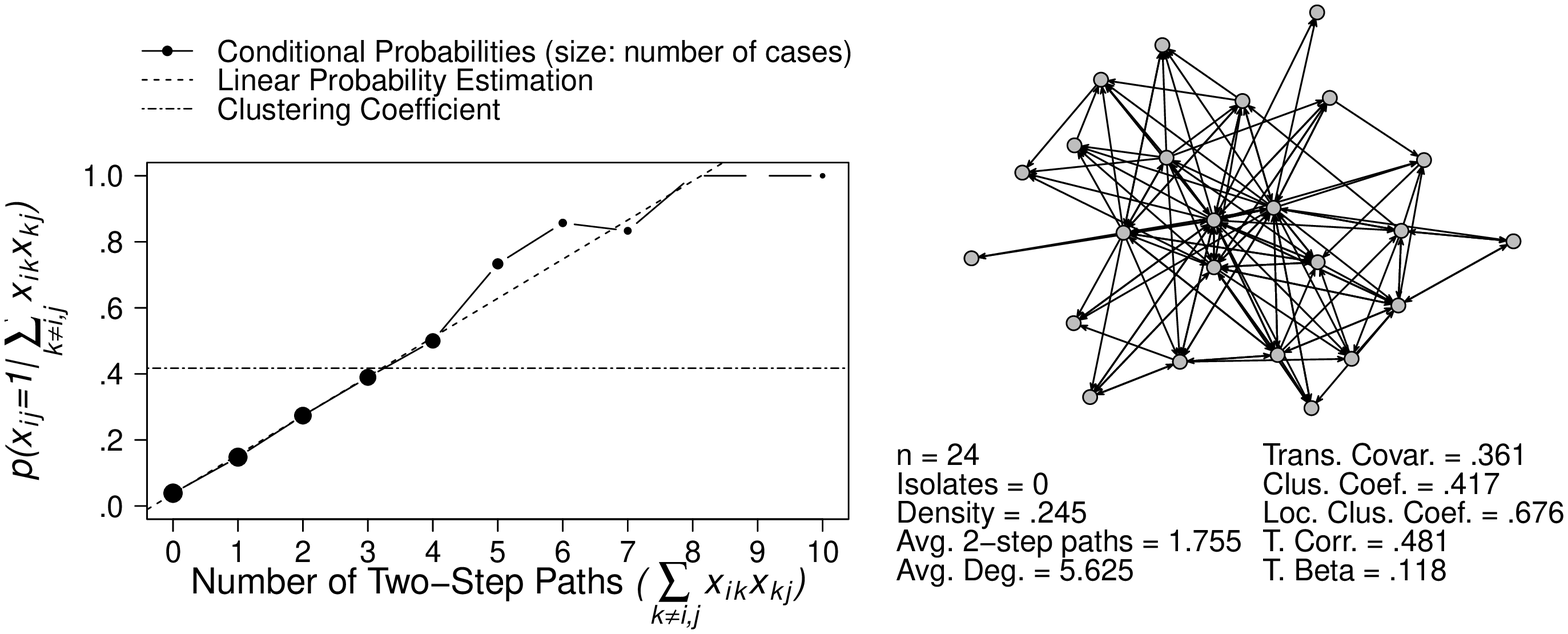}
	\caption{"24 Country Minerals and Fuels Trade"-Data \citep{SmithWhite1992, WassermanFaust1994}, cf. footnote \ref{fnmorenodata}.   }\label{fig:minerals}
	\end{subfigure}
\end{minipage}

\medspace

\begin{minipage}{1\textwidth}
		\begin{subfigure}[b]{.5\textwidth}
		    \includegraphics[width=1\textwidth]{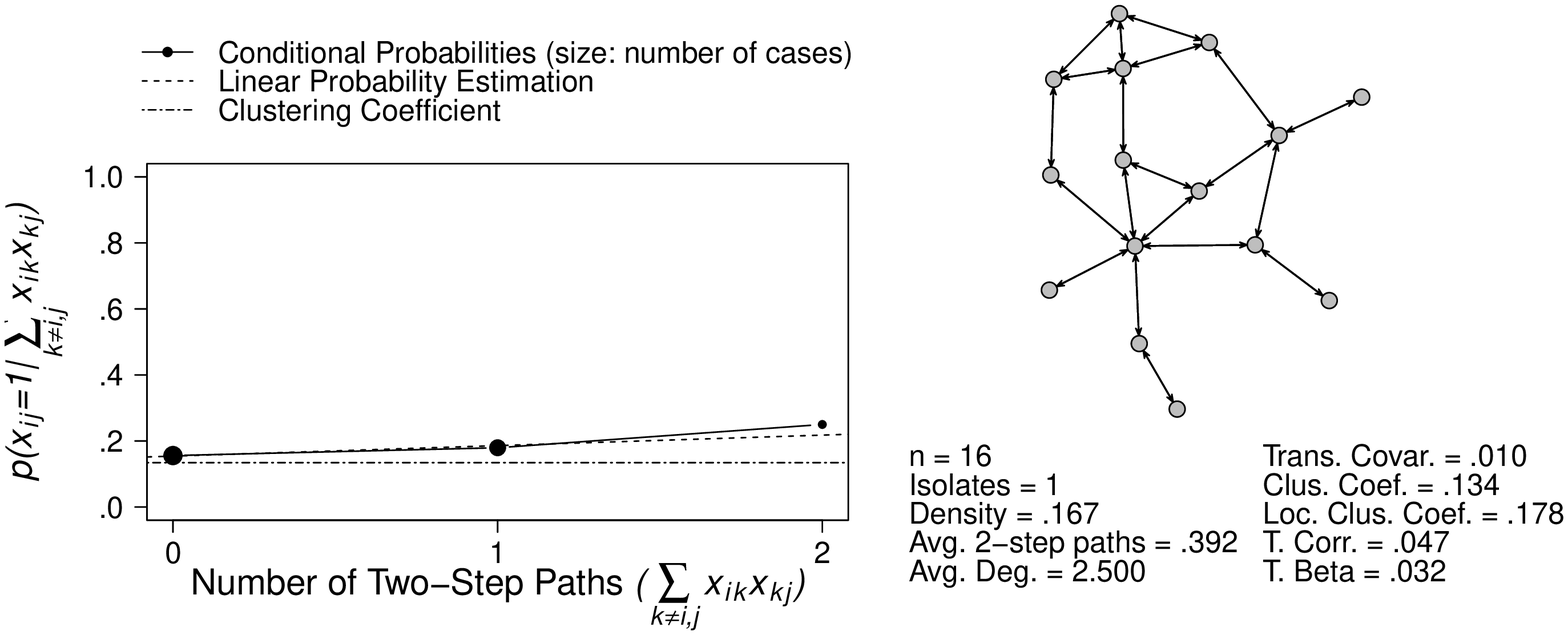}
	\caption{"Florentine Marriage"-Data \citep{PadgettAnsell1993}, cf. footnote \ref{fnmorenodata}.}\label{fig:marriage}
	    \end{subfigure}\quad
		\begin{subfigure}[b]{.5\textwidth}
			\includegraphics[width=1\textwidth]{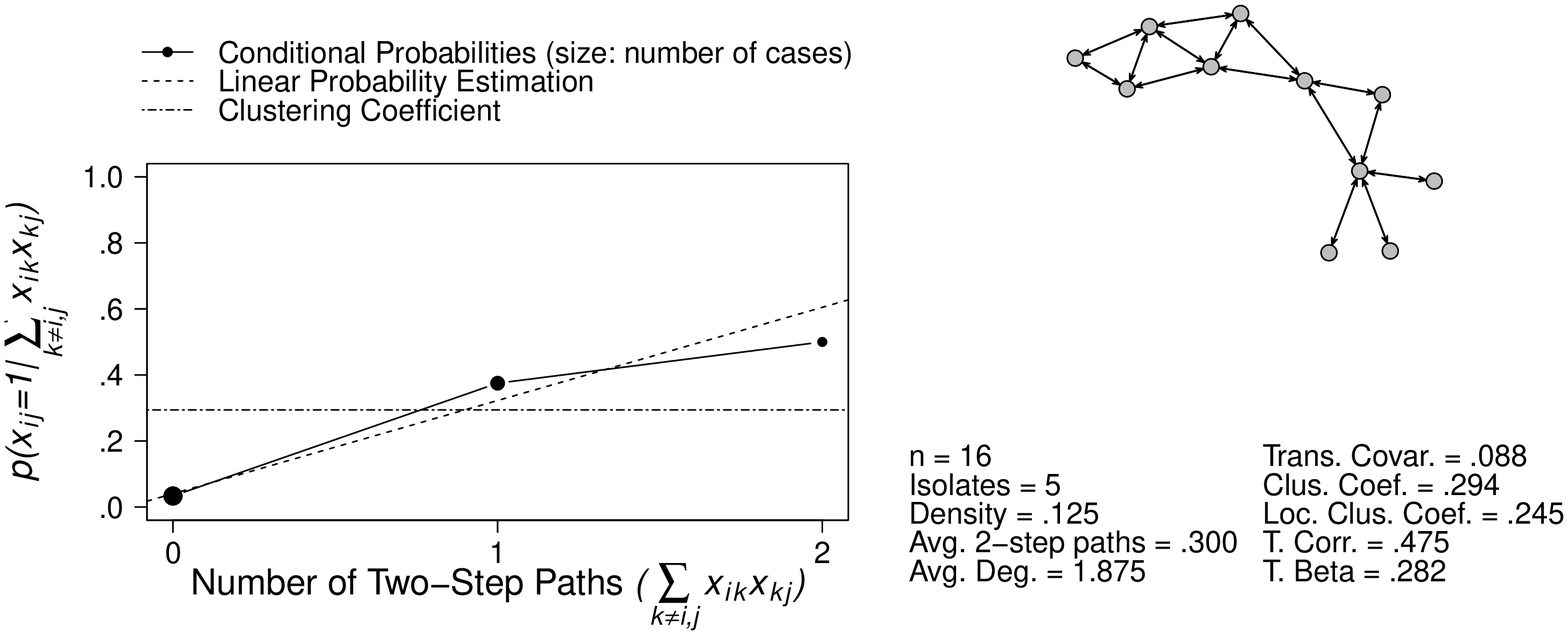}
			\caption{"Florentine Business"-Data \citep{PadgettAnsell1993}, cf. footnote \ref{fnmorenodata}.}\label{fig:business}
		\end{subfigure}
\end{minipage}
	\caption{C}

	\label{fig:empexa_C}
\end{sidewaysfigure}

In Figure (\ref{fig:empexa_A}: A, B, C), $12$ network datasets from different fields are analyzed. Each figure contains a diagram, an associated graph plot, and relevant summary statistics. The diagram shows the conditional probability for a tie given the number of two-paths on the vertical axis, and the number of two-paths on the horizontal axis. Information in the diagram is based on the depicted network although for clarity isolate nodes have been excluded.

Number of observations (ordered pairs of nodes) in each category of two-path counts is indicated by the size of dots. Each dot is connected with a straight line to emphasize the differences and direction in change of conditional probabilities between categories.

The horizontal dotted line indicates the clustering coefficient ($C$) for that network. This can be interpreted as the 'mean conditional probability' over all categories of two-path counts. By definition this measure discards all information about the differences between categories of two-path counts.

The dashed linear regression line between ties and number of ordered two-paths gives a linear approximation for these differences. The slope of this line is given by \textit{TB}, which hence allows for a network level indication of an increased (decreased) propensity towards transitivity. A down-side is that \textit{TB} doesn't allow for comparison between networks or a direct interpretation. However, \textit{TC} is a linear transformation of \textit{TB}, which serves these purposes.

The relevant summary statistics here are $n$, the number of nodes in the network, number of isolates (not depicted in the digraph plot), density, and the mean number of 2-step paths between the $n\,(n-1)$ node pairs, average degree (Avg.~Deg.), Transitivity covariance (Trans.~Cov.), Clustering Coefficient, \textit{C} (Clus.~Coef.), Local Clustering Coefficient, \textit{LC} (Loc.~Clus.~Coef.), transitivity correlation, \textit{TC} (T.~Corr.), and transitivity beta, \textit{TB} (T.~Beta).

The example networks are from a variety of fields, and differ in size ($\textit{n}=16$ to $\textit{n}=2114$) and structure ($\textit{d}=.001$ to $\textit{d}=.908$). In most examples there is a positive \textit{TB}, implying that in all these networks there is network transitivity. The exception is the formal organizational 'reports to' relationship among high-tech managers (Figure \ref{fig:reportsto}). The negative value for network transitivity here is induced by the design of formal organizational networks, which are usually set up as trees. Although in some examples a low clustering coefficient ($\textit{C}<.2$), such as for C.~elegans (Fig.\ \ref{fig:celegans}), protein interactions (Fig.\ \ref{fig:proteins}), and Mediaeval Florentine Family Weddings (Fig.\ \ref{fig:marriage}), could be interpreted as no tendency towards transitivity in the network, this would be a mistake. The positive regression coefficient \textit{TB} indicates an elevated propensity towards transitive triples occuring on average throughout these networks as the number of two-paths between pairs increases.

It must be emphasized that no inferential claims can be made about the statistical significance of these descriptive statistics. This would require further non-trivial assumptions about underlying digraph distributions. What could be done is to make a case by case comparison. For example, in the Florentine families data (Fig.'s \ref{fig:business} and \ref{fig:marriage}) it would be a valid statement to say that network transitivity is higher in the observed business network compared to the marriage network.

Further, this is not restricted to comparing networks on the same group of nodes, but holds for comparison between any type of network if we would compare \textit{TC}. For example, comparing the Southern women club with friendships in a law firm, the latter has (slightly) lower \textit{TC} ($.443$ vs.~$.445$), and hence lower network transitivity. In this case the clustering coefficient would have led to the same conclusion. But, this is not always so.

Comparing the inter country trade of minerals and fuel data (Fig.\ \ref{fig:minerals}) with frequent, and, very frequent information exchange (Fig.\ \ref{fig:RDadvice}) shows very similar diagrams. However, the clustering coefficients ($\textit{C}=.417$ and $\textit{C}=.344$, respectively) would suggest a different conclusion than when comparison is done on transitivity correlation ($\textit{TC}=.481$ and $\textit{TC}=.584$, respectively). This is due to differences in density of the two networks. The conditional probability of a transitive triple is higher in observed mineral and fuel trade network compared to information exchange, due to a higher density. The increased propensity towards transitive triples is \textit{more increased} in the information exchange network, and in this sense it shows more network transitivity.

Further, a remarkable finding that illustrates the value of these plots is that in three cases (Fig.'s \ref{fig:celegans}, \ref{fig:Rodriguez}, \ref{fig:proteins}) with positive \textit{TB} the probability of a tie doesn't show a monotonic increase with increasing two-path counts. Most clearly this is shown in the neural network of C. elegans, where beyond $9$ two-paths between two nodes, the probability for a tie strongly diminishes (except at $14$ two-paths). Reasons for this could be myriad, but it is important to consider that it could be indicative of missing, incomplete or biased data. The example in Figure (\ref{fig:proteins}) has been shown to be an incomplete dataset, which limited conclusions of the study on this dataset \citep[see for critiques][]{Coulombetal2005,Hanetal2005,Stumpfetal2005}. Or, due to ill defined relationships, for example, interactions could traverse through different media not considered (e.g., complementary use of email and phone), so that not all relevant interactions may have been observed. Similarly, the network in Figure (\ref{fig:Rodriguez}) displays a drop in tie probability at $3$ and $6$ two-paths, while a sharp increase occurs at $7$. As this dataset is a covert network constructed from secondary sources it could be indicative of a missing source, or a bias because some sources are irrelevant or receive too much emphasis. At least, non-monotonicity in the plots deserves a further theoretical explanation when no data related reasons can be found.

\section{Discussion}

This paper proposed new measures for transitivity based on covariances and correlations between ties and two-paths, and described some of their numerical properties. This new set of measures all are expressions of the difference in conditional probabilities in (\ref{bal1prob}) that define network transitivity.

\subsection{Statistical Inference}

The measures are proposed as descriptives, and not primarily for use in statistical inference. (For an overview of issues in statistical modeling for social network analysis see Snijders, 2011.) \nocite{Snijders2011}
Statistical inference about transitivity in networks can be directed at testing the null hypothesis of no transitivity, or at making statistical network models that do include transitivity. 
The former topic is treated by \citet{Karlberg1999}. This author defines two transitivity indices as potential test statistics, and uses as a null distribution the  \textit{U $\mid$ (OD, ID)} specification, i.e. the uniform distribution conditional on given in- and out-degree vectors. His first test statistic is (\ref{ClusCoef}).
His second test statistic is an average of local transitivity indices, where the local transitivity is defined as the density of the out-neighborhood of the node, divided by the maximal density given the indegree, outdegree, and number of mutual ties of the node. This reflects the importance of accounting for outdegrees, indegrees, and number of mutual ties that we encountered in (\ref{TC0}), the condition for the Transitivity Correlation to be 0. We suggest that our proposed statistic \textit{TC} could also be a suitable statistic for testing transitivity, and a suitable null distribution could be \textit{U $\mid$ (OD, ID, M)}, the uniform distribution  conditional on given in- and out-degree vectors and a given number $M$ of reciprocated ties.
(Note that \textit{OD} and \textit{ID} imply the values of $n$ and $d$.)
 Although generating random networks from these distributions is not discussed here, it should be noted that generating samples from the
 \textit{U $\mid$ (OD, ID)} as well as from the \textit{U $\mid$ (OD, ID, M)} distribution faces  serious combinatorial restrictions.
A computer program that can simulate samples from these two distributions
is ZO \citep{SnijdersZO}, based on \citet{Snijders1991},
and obtainable from \url{http://www.stats.ox.ac.uk/~snijders/socnet.htm}.
More recently a method for doing this was proposed by \citet{Tao2016}.
Further literature about the generation of networks with given in- and
out-degrees is \citet{Rao1996}, \citet{RobertsJM2000}, \citet{Verhelst2008},
and \citet{ChatterjeeDS2011}.

\subsection{Absence and Presence of Transitivity}

One of our conclusions is that condition (\ref{TC0}), depending on outdegrees, indegrees, and number of mutual ties, expresses absence of transitivity. This echoes and refines Feld and Elmore's \citeyearpar{Feld1982} observation, extended later by \citet{Faust2007}, that interpretations of the number of transitive triplets in a network should take into account the degree distributions. It is also related to the statement, made by \cite{Snijdersetal2006} and \citet[][p.\ 70]{Lusher2012}, that the number of independent two-paths (also called dyadwise shared partners) should be included in specifications of Exponential Random Graph Models as a `prerequisite', or lower-order configuration, for testing the transitive closure expressed by $k$-triangles (also called edgewise shared partners).

In the observed network examples in Section~\ref{ObsNet} we have mainly found positive values for Transitivity Covariance. This unambiguously shows that there is an increased propensity towards transitive triplets in these networks, in line with the predominance of transitive triplets found in a much larger set of networks already by \citet{Davis70}. However, in some cases the diagrams that depict the slope \textit{TB}, also show that the observed probabilities for ties may become highly variable for high values of the number of two-paths. This in itself is thought-provoking theoretically, and might inspire other measures that express deviations from a linear relation. However, other explanations are also possible, such as randomness, lack of data quality, or existence of covert ties.

\subsection{Extensions}

Next to transitivity we may consider balance \citep{Heider1958}. When  balance is treated for graphs or digraphs without considering edge signs, it is usual to treat absent edges as negative ties. Instead of Transitivity Covariance, the 'Balance Covariance' would then be based on the association between $x_{ij}$ and $x_{ik}\,x_{kj} + x^c_{ik}\,x^c_{kj}$, where $x^c$ is the complement of digraph $x$, with tie variables $x^c_{ij}=1-x_{ij}$. As the values of $x_{ik}\,x_{kj} + x^c_{ik}\,x^c_{kj}$ still are in $\{0,1\}$ the analyses will remain similar. The measure can then be further adjusted to accommodate other statements about triads.

Further refinements could be made regarding, for example, the implicit assumptions about homogeneity of nodes. In case nodes are explicitly organized in groups a distinction between different subsets of nodes, or different blocks of ties, may refine conclusions about increased, or decreased, levels of a tendency towards transitivity. Adjusted covariance based measures could be derived in this way, controlling for grouping of nodes.

Further developments could also be made for networks with valued ties. A generalized form of network transitivity for valued ties was proposed by \citet{OpsahlPanzarasa2009}. It is still unknown in which way this would lead to different conclusions and interpretations than those presented here.

\section{Conclusion}

We defined two new measures for transitivity: Transitivity Phi \textit{TPhi}, defined as the observed correlation between the tie variable between two nodes and a random two-path connection between them; and the Transitivity Correlation \textit{TC}, the observed correlation between the tie variable and the number of two-paths between the two nodes. The foremost advantage of these measures is that they offer a quantitative expression for the '\textit{increased} propensity' of transitive triples which is the definition of transitivity as formulated, e.g., by \citet{Newman2001}. By contrast, the clustering coefficient \textit{C}, one of the basic measures for transitivity, reflects the observed conditional probability of a tie, given a two-path, not a comparative quantity. Under the Erd\"os-Renyi model the clustering coefficient can have any expected value in $(0,1)$ depending on the density.
Because of their comparative nature these correlation measures allow for comparison between networks, even networks of unequal size or density, and from different contexts.

The two measures are both based on considering the tie variable for a random pair $(i,j)$ of nodes; the difference is that \textit{TPhi} considers one randomly selected third node, whereas \textit{TC} considers all other nodes as potential intermediates. Both are functions of the ego-networks of all nodes in the digraph, where the ego-network is defined as the digraph induces by the node and all nodes in its direct out-neighborhood. Clearly, \textit{TC} takes into account much more of the structure of the ego-networks than \textit{TPhi}, specifically, the dependence between the different two-paths connecting any two nodes.

The results found in the comparison of measures for windmill graphs led to the conclusion that the difference between these two measures can imply large differences in conclusions about transitivity. For windmills with many wings the consideration of the  two-path dependence by \textit{TC} leads to a value tending to 1, contrasting with the value for \textit{TPhi} tending to 0. We interpret windmill graphs as being highly transitive, and find this a strong argument in favor of \textit{TC} over \textit{TPhi}.

Correlations between binary variables are known to have a restricted range.
For graphs that are unions of disconnected complete subgraphs of equal sizes, both \textit{TC} and \textit{TPhi} assume the maximum of~1. 
This shows that there may be room for developing other measures for transitivity that assume their maximum value for all totally transitive graphs, without the restriction of equal-size components.

A finding that we believe to be new is that the condition that \textit{TC} is zero, is equivalent to a condition on the covariance between in- and outdegrees, the number of mutual ties, the density, and the number of nodes. This leads to interest in the uniform distribution for digraphs conditional on these four quantities. This distribution presumably is very difficult to handle; the distribution of digraphs, for a given number of nodes, conditional on the vectors of in- and outdegrees and the number of mutual ties may be presumed to be easier to handle, although this distribution already poses huge problems \citep{Tao2016,SnijdersZO}.

\bibliography{TCbib}
\bibliographystyle{nws}

\begin{appendices}
	\section{Relation \textit{TPhi} and \textit{TC}}\label{AppTPhiTC}
	
	The covariance between directed ties and the number of two-paths is,
	
	\begin{equation}\label{covsumS}
	\text{cov}\Big(x_{ij},\sum\limits_{h\neq i,j}x_{ih} x_{hj}\Big)=\frac{1}{n\,(n-1)}  \sum_i\sum_{j\neq i}x_{ij} \,\big(\sum_{h\neq i,j}x_{ih}\, x_{hj}\big) - \myovline{x}\,\myovline{xx},
	\end{equation}
The difference between \textit{TC} and \textit{TPhi} is in scaling. Consider that covariance in (\ref{covsumS}) is a weighted measure of the numerator in (\ref{covPhi}),
	\begin{equation}\label{covTCTPhi}
	\text{cov}\Big(x_{ij}, \sum_{h\neq i,j} x_{ih}x_{hj}\Big) = (n-2)\, \text{cov}(x_{ij},\, x_{ik} x_{kj}).
	\end{equation}
	Further, the denominator in (\ref{TC}) differs from that of \textit{TPhi} only in $s.e.(\sum_{h}x_{ih}x_{hj})$ (see (\ref{TPhi})). The variance of the number of two-paths between any ordered pair can be rewritten as
	\begin{align}\label{varTS}
	\text{var}\Big(\sum_{h}x_{ih}x_{hj}\Big)=
           (n-2)\,\big[\text{var}(x_{ik}x_{kj}) + (n-3)\, \text{cov}(x_{ik}x_{kj},x_{i\ell}x_{\ell j})\big]
	\end{align}
	Under conditions where $\text{cov}(x_{ik}x_{kj},x_{i\ell}x_{\ell j})=\text{var}(x_{ik}x_{kj})$, \textit{TC} reduces to \textit{TPhi} as
	\begin{align}\label{varTS2}
	\text{var}\Big(\sum_{h}x_{ih}x_{hj}\Big)= (n-2)^2 \, \text{var}(x_{ik}x_{kj}).
	\end{align}
	But, more generally, we can state
	\begin{align}\label{eq:TC=TP}
	\textit{TC}&= \alpha \times \textit{TPhi},
	\end{align} where $\alpha$ is
	\begin{align}\label{eq:A}
	\alpha&=\frac{\textit{TC}}{\textit{TPhi}} \nonumber \\ &=\frac{(n-2)\sqrt{\text{var}(x_{ik}x_{kj})}}
{\sqrt{((n-2)[\text{var}(x_{ik}x_{kj})+(n-3)\, \text{cov}(x_{ik}x_{kj},x_{i\ell}x_{\ell j})])}}
	\end{align}
	for $\ell \neq k$. Note that (\ref{eq:A}) can be rewritten as
	
	\begin{align}\label{eq:A2}
	\alpha= \frac{(n-2)}{\sqrt{((n-2)\, [1+(n-3)\rho])}},
	\end{align}
	where
	\begin{align}\label{eq:Arho}
	\rho = \frac{\text{cov}(x_{ik}x_{kj},\, x_{i\ell}x_{\ell j})}{\text{var}(x_{ik}x_{kj})} 
	\end{align}
 is the autocorrelation between two-paths in a digraph. 
 Now $\rho$ is a correlation so that $\rho \leq 1$; further, rewriting (\ref{varTS}),
	\begin{align}\label{eq:A3}
	0 \,& \leq \,  \text{var}\Big( \sum_{k \neq i,j} x_{ik}x_{kj}\Big) \,=\,
           (n-2)\, \text{var}(x_{ik}x_{kj}) \,+\, (n-2)(n-3)\,\text{cov}(x_{ik}x_{kj}) \\
            & \,=\,  (n-2)\big( 1 + (n-3)\,\rho\big)\, \text{var}(x_{ik}x_{kj}) \ , \nonumber
	\end{align}
 which implies that $\rho \geq -1/(n-3)$. With (\ref{eq:A2}) this implies that
 $\alpha \geq 1$ except for $\textit{TPhi}=0$, where $\rho = -1/(n-3)$ and $\textit{TPhi}$ is undefined. It can be concluded that $\textit{TC} \geq \textit{TPhi}$. The distinction between \textit{TC} and \textit{TPhi} is about size not direction. Although the relation between $\alpha$ and $\rho$ is non-linear, $\alpha$ is monotonically decreasing as $\rho$ increases. The ratio of the two transitivity correlations is a function of $n$ and the two-path autocorrelation in the digraph.
	The autocorrelation between two-paths is itself a Phi-coefficient, expressing the difference in conditional probabilities of a two-path via a node $k$ given a two-path via another node $\ell$ exists and of a two-path via $k$ given that no other two-path exists. As such, it can be interpreted as a measure of network centrality, where a smaller $\rho$ indicates an elevated uniqueness of nodes as intermediate in two-paths.
	
	\section{When is Transitivity Covariance Equal to Zero} \label{app:Properties TCov}

We derive an condition equivalent to the property that the Transitivity Correlation $\textit{TC}$, or equivalently the Transitivity Covariance, is zero.

	The Transitivity Covariance is defined as the covariance, for a randomly chosen pair $(i,j)$, between the direct tie $x_{ij}$ and the number $\sum_{h\ne i,j} x_{ih}x_{hj}$ of directed two-paths between these nodes as defined in (\ref{covTCTPhi}).
	Network density in (di)graphs is the mean tie indicator variable,
	
	\begin{equation}\label{dx}
	d=\myovline{x}=\frac{1}{n(n-1)} \sum_{i=1}^n \,\, \sum_{\mathclap{\substack{j=1\\ j\neq i}}}^n \, x_{ij} \ .
	\end{equation}
	Over all ordered pairs $(i,j)$, the mean number of two-paths is	
	\begin{equation}\label{dx2}
	\myovline{xx}=TSP= \frac{1}{n(n-1)} \sum_{i=1}^n \,\, \sum_{\mathclap{\substack{j=1 \\j\neq i}}}^n \enskip \, \sum_{\mathclap{\substack{h=1 \\h\neq i,j}}}^n \, x_{ih}\, x_{hj} 
	\end{equation}
	and the mean number of transitive triples  is given by
	\begin{equation}\label{dxx2}
	TT= \frac{1}{n(n-1)} \sum_{i=1}^n \,\, \sum_{\mathclap{\substack{j=1 \\j\neq i}}}^n \enskip \, \sum _{\mathclap{\substack{h=1 \\h\neq i,j}}}^n \, x_{ij}\, x_{ih}\, x_{hj} \ . 
	\end{equation}
	Let
	
	\begin{equation}\label{Mutuals}
	M=   \sum_{i=1}^n \, \sum_{\mathclap{\substack{j=1\\ j\neq i}}}^n  \, x_{ij} \, x_{ji}
	\end{equation}
	be the sum of reciprocal or mutual ties (note that reciprocal ties are two-paths that do not contribute to a transitive triple). 
	The mean number of two-paths in (\ref{dx2}) can be rewritten as
	
	\begin{equation}\label{cnt2steps}
	TSP =  \frac{1}{n\,(n-1)}(OD\cdot ID - M)
	\end{equation}
	where $ID$ and $OD$ are the vectors of indegrees and outdegree, and $OD\cdot ID$ is their inner product.
	
	Substitution in (\ref{covTCTPhi}) gives
	\begin{equation}
	\text{cov}(x_{ij},\sum_{h \neq i,j} x_{ih} \, x_{hj}) = (n-2)\Bigg(TT-\frac{(OD\cdot ID -M) \, d }{n (n-1)}\Bigg) \ .
	\end{equation}
	
	The inner product of two vectors can be expressed in terms of covariance, in this case the covariance between indegree and outdegree for the probability distribution that a node is randomly chosen. This gives
	\begin{equation}
	\text{cov}(OD,ID)= \frac{1}{n}(OD\cdot ID) - d^2 \ .
	\end{equation}
	Substitution gives

	\begin{equation}
	\text{cov}(x_{ij},\sum_{h \neq i,j}^n x_{ih}\, x_{hj}) = (n-2) \Bigg(TT-\frac{\big(n  \,\, \text{cov}(OD,ID) +n \, d^2 -M\big)\, d }{n (n-1)}\Bigg),
	\end{equation}
which simplifies to equation (\ref{TC0}) under the condition of no network (in)transitivity.	
\end{appendices}

\end{document}